%% file: main.tex
\documentclass[a4paper,fleqn]{cas-sc}

\usepackage{float}
\usepackage{placeins} % \FloatBarrier

\usepackage[english]{babel}
\usepackage[round]{natbib}
\usepackage[utf8]{inputenc}

\usepackage{bm, bbm}
\usepackage{graphicx}

\usepackage{import} % for importing SI

\usepackage{xurl}

\usepackage[nottoc,numbib]{tocbibind} %to make references appear in table of contents

\usepackage{todonotes}
\usepackage{multirow}

\usepackage{color}
\usepackage{soul}

\AtBeginDocument{%
 \hypersetup{
  citecolor=Bittersweet,
    linkcolor=Blue,
  urlcolor=Blue}
}

\makeatletter
\def\ps@first{%
   \let\@oddhead\@empty
   \let\@evenhead\@empty
   \def\@oddfoot{}
   \let\@evenfoot\@oddfoot
}
\ExplSyntaxOn
\cs_gset:Npn \__first_footerline:
  { \group_begin: \small \sffamily \__short_authors: \group_end: }
\ExplSyntaxOff

\begin{document}

\title[mode = title]{
Economic impacts of a drastic gas supply shock and short-term mitigation strategies
}

\shortauthors{Pichler, Hurt, Reisch, Stangl, Thurner}
\shorttitle{Economic impacts of a drastic gas supply shock and short-term mitigation strategies}

\author[1,2]{Anton Pichler}[orcid=0000-0002-7522-1532]
\cormark[1]
\author[2]{Jan Hurt}%[orcid=xxx]
\author[2]{Tobias Reisch}[orcid=0000-0002-1219-2893]
\author[2]{Johannes Stangl}%[orcid=xxx]
\author[2,3,4,5]{Stefan Thurner}%[orcid=xxx]

\address[1]{Vienna University of Economics and Business, Austria}
\address[2]{Complexity Science Hub Vienna, Austria}
\address[3]{Section for Complex Systems, Medical University of Vienna, Austria}
\address[4]{Supply Chain Intelligence Institute Austria, Austria}
\address[5]{Santa Fe Institute, USA}

\cortext[cor1]{Corresponding author: anton.pichler@wu.ac.at}

\date{\today}

\begin{abstract}
\noindent 
The Russian invasion of Ukraine on February 24, 2022 
entailed the threat of a drastic and sudden reduction of natural gas supply to the European Union.
This paper presents a techno-economic analysis of the consequences of a sudden gas supply shock to Austria, one of the most dependent countries on imports of Russian gas. 
Our analysis comprises (a) a detailed assessment of supply and demand side countermeasures to mitigate the immediate shortfall in Russian gas imports,
(b) a mapping of the net reduction in gas supply to industrial sectors to quantify direct economic shocks and expected relative reductions in gross output and
(c) the quantification of higher-order economic impacts through using a dynamic out-of-equilibrium input-output model. 
Our results show that potential economic consequences can range from relatively mild to highly severe, depending on the implementation and success of counteracting mitigation measures. We find that securing alternative gas imports, storage management, and incentivizing fuel switching represent the most important short-term policy levers to mitigate the adverse impacts of a sudden import stop.
\end{abstract}

\begin{keywords}
Gas dependency \sep Energy security \sep Energy policy \sep Production networks \sep Shock propagation \\[.75em]
\end{keywords}

\maketitle
%\tableofcontents

%%%%%%%%%%%%%%%%%%%%%%%%%%%%%%%%%%%%%%%%%%%%%%%%
\section{Introduction}
\label{sec:intro}
\FloatBarrier

\begin{figure}%[ht]
    \centering
    \includegraphics[width=0.9\textwidth]{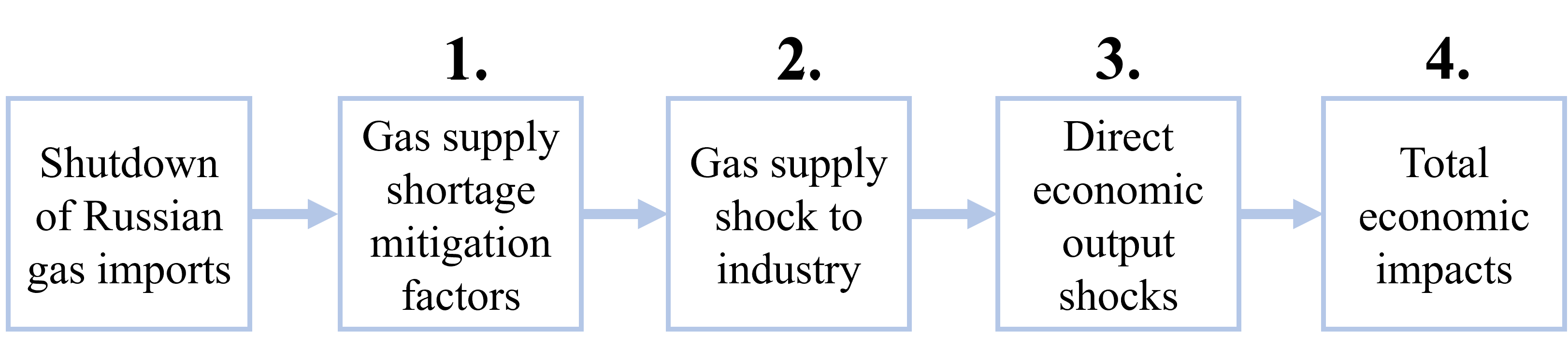}
    \caption{ 
    {\bf Outline of our analytical approach.}
    We consider a complete hypothetical shutdown of Russian gas imports to Austria as of June 1, 2022. As first step, we analyze factors on the supply and demand side that have the potential to mitigate the immediate shortfall in gas supply. Second, after taking mitigation factors into account, we quantify gas supply shocks to industrial production. Third, we estimate direct production losses caused by gas shortages. Fourth, using a dynamic input-output model, we quantify total (direct plus indirect) impacts on the economy.
    }
    \label{fig:fig1}
\end{figure}

The intensification of the Russo--Ukrainian conflict has plunged Europe into the worst energy crisis since the 1970s.
In 2021, the European Union (EU) sourced about 38\% of its annual gas consumption from Russia (see Supplementary Information [SI]) but experienced a significant decline in its import stream over the year. With the Russian invasion of Ukraine on February 24, 2022, the possibility of a complete halt of Russian gas imports to the EU became a highly realistic scenario. Such a scenario could have materialized through an EU embargo on Russian energy products or an export stop enforced by Russia.

We focus on the potential economic impacts in Austria resulting from a complete halt of natural gas imports from Russia. Austria represents a particularly interesting study case. As of 2021, about 80\% of Austria's gas consumption has been imported from Russia, making it one of the most exposed countries to imports of Russian gas. However, Austria is a relatively small country, representing only about 2\% of the EU-wide gas consumption (see SI for computations), whose economy is strongly integrated into the European single market. Our paper investigates how these factors influence the severity of the economic impact of a drastic gas supply shock.

It is important to note that the results presented here have been developed in a prospective study that we have published as a Policy Brief on May 24, 2022 \citep{pichler2022austria}. 
Here, we provide additional results and put our analysis into the context of other studies and discuss what we have learned with the help of hindsight. 
Our baseline scenario considers a hypothetical complete halt of Russian gas imports to Europe starting on 1 June 2022, resulting in a reduction of 80\% of Austria's gas supply in terms of annual consumption. In such a scenario, the national Energy Intervention Powers Act (``Energielenkungsgesetz'') \citep{ris22_intervention} would likely become relevant, granting policymakers a key role with the possibility to intervene in energy markets to secure energy supplies to critical customers. Here, we assess the potential outcomes of various possible policy actions.

Our approach combines technical and economic aspects of the recent episode of Austria's gas crisis and is summarized visually in Fig~\ref{fig:fig1}.
As a first step, we analyze demand and supply side factors that have the potential to mitigate the adverse consequences of an immediate import shock. More specifically, we consider the mitigation potential of alternative gas imports, gas storage management, fuel switching, and savings in heating and pipeline operation. We use these factors to derive two illustrative gas shock scenarios. First, Scenario A assumes a coordinated distribution of scarce gas supplies among EU member states to mitigate economic impacts for disproportionately exposed countries. Second, Scenario B assumes no coordination. Instead, the countries access the reduced European supply based on their relative market share.
As a second step, we quantify the amount of gas reduction to economic production after taking into account customer prioritization that would likely happen if the \textit{Emergency level} of the national Gas Emergency Plan is reached \citep{bmk22_emergency}. As a third step, we map these gas supply reductions to individual industrial sectors and estimate their immediate impacts on sectoral outputs. 
In a final step, we initialize a dynamic disequilibrium input-output model introduced in \cite{pichler2022forecasting} to assess the total (direct + indirect) impacts on the Austrian economy. In particular, we use the economic model to estimate total aggregate impacts on output, profits, wages and salaries, and private consumption and investigate sectoral impacts. 
It is important to note that any analysis of the economic impacts of a drastic gas supply reduction involves a high level of uncertainty. Consequently, we focus only on the short-term effects of a drastic gas supply shock and simulate the adverse economic impacts over a 2-month horizon in the immediate aftermath of the gas supply reduction.

Our approach integrates the different policy-relevant mitigation factors with an economic model, allowing us to assess the relative effectiveness of specific policy measures. Our results show that the expected gas reductions to industry are highly different for the two main scenarios considered. As a small, open, and landlocked economy, Austria would disproportionately benefit from EU-coordinated efforts to secure alternative gas supplies. In this scenario (Scenario A), we estimate a country-wide gas supply shock of about 5\% after mitigation measures have been considered, translating into a 10\% reduction of gas supply for economic production. In Scenario B, where we assume no coordination among EU-member states, we estimate a country-wide gas supply shock of about 24\%, resulting in a 53\% reduction of gas supply for economic production. Consequently, economic impacts vary substantially across the two scenarios. For the cooperative Scenario A and the uncoordinated Scenario B, we estimate that direct output shocks due to the gas supply shock amount to 1.1\% and 5.6\%, respectively. When feeding these direct output shocks into our sectoral economic model, we find amplification effects of about 1.5, resulting in short-term total economic impacts of up to -1.7\% and -8.4\% for Scenario A and Scenario B, respectively.

While these results are based on two specific, illustrative scenarios, we use our economic model to simulate the consequences of the whole spectrum of possible gas shocks. Our results show that for a plausible range of gas supply shortages, economic production is quite sensitive with respect to further incremental reductions in gas supply. In this range, we observe that a reduction of another 1\% of gas inputs reduces economic output by about 0.25\%, indicating that even policies leading to small improvements in gas availability during times of crisis can have substantial positive effects on the economy. 

These results also emphasize the large uncertainties surrounding any study on the economic impacts of a drastic gas import shock. In addition to inherent uncertainties of any economic model, a major source of uncertainty originates from the highly contingent realization of potential mitigating factors, ranging from access to alternative gas import sources to temperature effects during the heating season. Not surprisingly, estimates of the economic impact of a sudden gas shock to Europe have varied substantially across studies. For example, studying the consequences of a sudden import stop of natural gas to Germany, \cite{krebs2022auswirkungen} finds potential consequences to gross national income amounting to about -8\% but also considers scenarios which lead to substantially less adverse economic impacts about -1.2\%. Also looking at the German case, \cite{bachmann2022if} find relatively small impacts on gross domestic product (GDP) of about -0.3\% and estimate that impacts of -2.3\% represent extremely adverse outcomes. (For context, we apply their economic model to Austria in the SI).
Estimates by \cite{bundesbank2022} and \cite{bayer2022stopp} found GDP impacts in the range of about -2\% to -4\%, while estimates by \cite{behringer2022ukraine} and a consortia of various economic research institutes \citep{gd2022pandemie} lie between -2.4\% and -6\%.
Differences in these estimates arise not only from different modeling assumptions but also from authors' expectations and assumptions on plausible shock and mitigation scenarios. For a brief overview of various economic assessments, including short scenario descriptions, see \cite{berger2022potential}.

Most studies have focused on energy or gas import shocks to Germany, which represented the largest single importer of Russian gas in Europe. However, it is worth pointing out that the situation in Germany has been quite different from that in Austria. 
While Austria roughly consumes a factor of ten less gas than Germany, it exhibited (and still exhibits) much higher dependencies on Russian gas. Consequently, a sudden halt in imports of Russian gas would disproportionately affect Austria.
However, given Austria's smaller size, its ability to mitigate such import stops would have relied more heavily on the responses at the European level. 
One study investigating the impacts of the Russo-Ukrainian conflict on Austria is \cite{koeppl22}, that focuses on an Austrian export stop and gas prices rather than on a Russian gas import stop. 
A study based on similar shock timing and model assumptions is \cite{schneider22effekte}, that found a reduction of -3.1\% on Austria's GDP for 2022 resulting from a gas supply reduction of 37\%.

Our work relates to studies that investigated the economic impact of gas supply shocks independent of the Russo-Ukrainian War. For example, \cite{alcaraz2017effect} investigated the impact of gas shortages in the Mexican economy between 2012 and 2013.
The study highlights that states afflicted by gas shortages witnessed significant declines in manufacturing output when compared to regions unaffected by such shortages.
Overall, they found that gas shortages resulted in a 0.3\%-lower annual GDP growth rate in Mexico.  It is crucial to note, however, that the gas shortages examined in this study were localized, temporally constrained, and considerably less severe than those considered in our analysis. \cite{gul2023energy} studied a recent episode in T\"urkiye, where gas consumption of gas-intensive industrial facilities was reduced by 40\% for ten days due to severe import reductions.
They found that industries with stronger reliance on natural gas experienced a 5.6\% lower production growth rate after the shock compared to industries with weaker reliance.

The paper is structured as follows. We develop our main gas shock scenarios and discuss potential mitigation strategies in Section~\ref{sec:gasscenario}. In Section~\ref{sec:drect}, we map estimated gas shocks to industrial sectors and quantify direct production shocks. Building on a dynamic input-output model, we present aggregate and sectoral impact estimates 
in Section~\ref{sec:impact}. We discuss our analysis in the context of what has happened empirically in Section~\ref{sec:discuss} and conclude in Section~\ref{sec:conclude}. 
%Reproduction files can be found online\footnote{
%\url{https://doi.org/10.5281/zenodo.13753134}.
%}.

\section{Gas shocks and mitigation strategies} \label{sec:gasscenario}
We consider a complete halt of Russian gas exports to the EU as of June 1, 2022, consequently, tightening the EU-wide gas supply. Fixing a date for computing the gas shock is important since consumption patterns and gas storage levels are non-constant over the year, resulting in different gas shortages for different dates (see Appendix~\ref{apx:mitigate} for details). 
For Austria, at that time, a sudden Russian gas import stop translates into an immediate 80\% shock in terms of its annual consumption.
While the complete import stop did not realize\footnote{ 
In fact, as of early 2024, Russian gas still represents by far the largest source of gas imports to Austria (see Section~\ref{sec:empirical}).
}, 
such drastic reductions in gas supply were considered very plausible in the immediate aftermath of the Russian invasion and Austrian policymakers were seeking a better understanding of the implied economic consequences and potential mitigation strategies.

We focus on two main scenarios: In an EU-wide cooperation scenario (Scenario A), additional natural gas resources at the EU level are distributed in a coordinated way to avoid excessively high import shocks for particularly exposed countries. In this scenario, also storage levels are cooperatively managed; every member state proportionally bears the same reductions in their gas consumption, even if not exposed to Russian gas imports. 
In an uncoordinated scenario (Scenario B), Austria mitigates the Russian import shock by buying gas from alternative sources without coordinating with other EU member states and managing its gas storage capacities individually.

We explain the details of the two scenarios in Appendix~\ref{apx:mitigate} and summarize its main pillars in Table~\ref{tab:scen}. For both scenarios, the 80\% reduction of gas supply does not translate one-to-one into a gas supply shock to the economy. 
Instead, there are several short-term measures that can be undertaken to mitigate the adverse impacts of a sudden gas supply import shock. Due to the sheer magnitude of the gas import shock, it is likely that in such a case, great effort will be undertaken to reduce the country's short-term gas dependency, even if it is costly. We consider various aspects that  impact the actual gas shock to the Austrian economy, including additional imports of liquified natural gas (LNG) and gas from alternative suppliers, storage management, and fuel switching in the power and heat sector. 

As outlined in Appendix~\ref{apx:_mit_imp}, both scenarios assume that the EU is able to secure additional imports of about 55 bcm, accounting for more than a third of previous gas imports from Russia. This assumption is slightly more conservative than the EU Commission’s strategy put forward in early 2022 \citep{eu22_repower}.
The main difference between the two scenarios is the amount of these additional EU-wide supplies that Austria can access. In the (optimistic) EU-wide cooperation scenario, Scenario A, the EU-wide gas supply shock is evenly distributed among member states. In this case, Austria is able to secure about 5.20 bcm (corresponding to 55.7\% of annual consumption) of additional gas supplies over the next year. In the (pessimistic) uncoordinated scenario, Scenario B, member states would access additional EU-wide supplies based on existing consumption shares. In this case, we estimate that only 2.65 bcm (28.4\%) could be imported additionally  (Appendix~\ref{apx:_mit_imp}).
Note that Austria exhibits extremely large import shares of Russian gas but represents only a small fraction of EU-wide gas consumption. Thus, Austria disproportionately benefits in Scenario A, where EU-wide gas import shocks are transmitted to member states without considering existing import shares. In Scenario B, however, where the reduced EU-wide gas supplies are distributed according to existing consumption shares, gas shocks to Austria are considerably higher.

Compared to its 2.3\% share of EU gas consumption, Austria's gas storage capacities are large, accounting for almost 9\% of total EU storage capacities. We assume that gas storage can be accessed to smooth out import reductions of natural gas and use a simple storage model to quantify the usage and availability of natural gas in storage facilities (Appendix~\ref{apx:_mit_store}). On June 1, 2022, when the hypothetical Russian import stop becomes effective, we estimate the total storage levels of the EU and Austria to be 58.02 bcm and 3.56 bcm, respectively.
In Scenario A, where EU member states cooperate in managing the complete halt of Russian gas, the disproportionately large storage facilities located in Austria will also be used to smooth out gas supply shortages in other EU countries. In this case, we estimate that Austria could access 0.64 bcm of the gas extracted from its storage (6.9\% of annual consumption). In an uncoordinated scenario, however, Austria could extract larger amounts from its storage facilities to mitigate gas supply reductions and estimate an extraction potential over the next year of  1.40 bcm (15\%). We discuss the usage of gas storage to mitigate in more detail in the Discussion section (Section~\ref{sec:conclude}). 

Considering a complete halt of Austria's imports of Russian gas, alternative import sources, and the extraction potential of existing gas storage capacities results in an aggregate gas supply shock of 1.63 bcm and 3.42 bcm for Scenario A and Scenario B, respectively. Besides alternative imports and storage management, the next most relevant mitigation factor is fuel switching, although there are substantial uncertainties around time constraints and the magnitude of such measures. As detailed in Appendix~\ref{apx:_mit_sub}, we assume the possibility of short-term fuel substitution of gas power plants providing electricity baseload\footnote{ \label{foot:switch}
Note that fuel substitution is likely to be costly due to both additional investment costs and increased operational costs. However, considering the massive threat to energy security due to a sudden stop of Russian gas imports, we consider the implementation of fuel substitution measures as very likely. Most likely, such measures would be mandated or at least financially supported by the government. In fact, in mid-June 2022, the Austrian government instructed the largest Austrian power provider, Verbund (which is majority-owned by the Republic of Austria), to put into operation an already decommissioned Coal plant in Mellach to produce electricity in case of gas shortages: 
\url{https://www.derstandard.at/story/2000136684272/anhaltend-reduzierte-gasfluesse-aus-russland-schueren-aengste} (accessed: 16 Jan 2024).
}. Overall, we estimate that almost 1 bcm of natural gas could be saved by switching to alternative fuels in electricity and heat production.

We also consider the potential of gas savings from reduced room temperatures in winter. We do not model the impact of increased gas prices on gas consumption for heating explicitly but assume a 1$^{\circ}$C reduction in average room temperature and are agnostic about its causes (e.g., price increases or public campaigns). We estimate that a reduction of 1$^{\circ}$C translates into savings of around 0.11 bcm in the residential sector, but discuss in Appendix~\ref{apx:_mit_heat} that these assumptions are rather conservative. In Section~\ref{sec:empirical}, we provide further context on how gas consumption patterns have changed in the past years. 
As a final point, note that gas usage for operating the pipeline infrastructure is non-negligible and depends on the amount of gas transported. Due to the lower gas consumption levels, gas usage for operating pipelines declines by 0.05 bcm and 0.11 bcm in Scenario A and B, respectively (Appendix~\ref{apx:_mit_pip}).

Taking all the short-term dampening effects into account, we obtain a reduction of gas availability for Austria between June 2022 and June 2023 of 0.49 bcm in Scenario A, and 2.21 bcm in Scenario B. Compared to 2021 consumption levels, this results in an aggregate gas supply shock of 5.2\% in the EU-cooperation scenario and 23.7\% in an uncoordinated scenario. 
To model the economic consequences of the aggregate gas shock, we have to determine how much of the remaining gas supply can be used for economic production.

In accordance with recent communications from the government
\citep{bmnt19, bmk22}, we assume that other consumer types, including households, electricity, and the public sector, are prioritized over industrial companies in case of gas shortages. 
To calculate the gas available for industry consumption, we first compute the consumption of protected consumers after all consumption-reducing factors have been considered. This comprises household consumption (after lowering room temperature), electricity generation (reduced by fuel switching), and consumption in pipelines (with reduced throughput). Second, we subtract this number from the gas supply to Austria in the respective scenario (A or B), resulting in the amount of gas that is available for industry. Since about 28\% of industry consumption is attributed to room heating, we also assume that industry reduces the average room temperature by 1$^{\circ}$C, resulting in a reduction of overall industry gas demand of 0.06 bcm (-0.61\%). In our model, we calculate the industry gas shock as the shortfall relative to the sum of (reduced) room heat, process heat, and non-energetic use, assuming that the relative allocation between the three usages stays constant.

\begin{figure}[!ht]
    \centering
        \includegraphics[width=\textwidth]{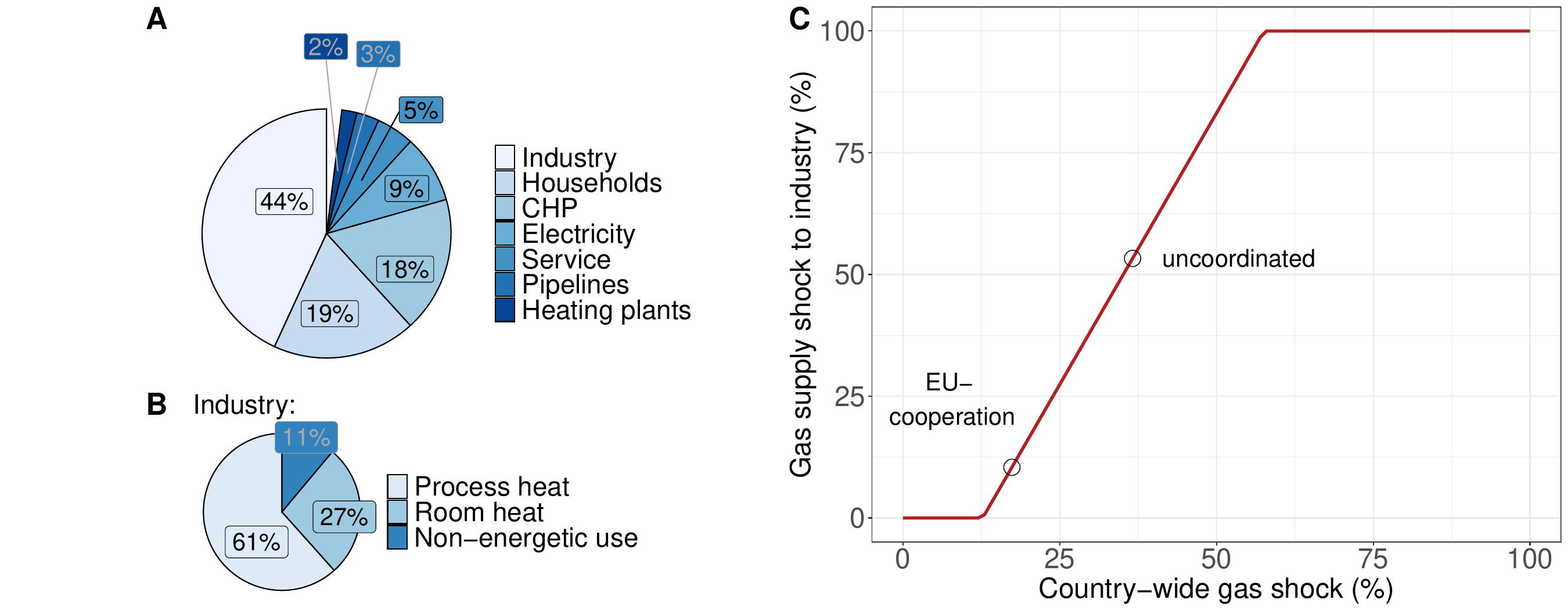}
    \caption{
    {\bf A: Relative gas consumption across consumer types.} 
    ``CHP'' stands for combined heat and power, ``Pipelines'' refers to gas distribution, and ``Service'' includes the service and the public sectors. Data is based on 2021 \citep{stat_nutzenergie}.
    {\bf B: Gas consumption across different industrial 
    %usage 
    categories.} 
    This pie chart zooms into industry-specific consumption. The room and process heat numbers are from \cite{stat_gesamtenergie}. 
    {\bf C: Mapping country-wide gas supply shocks to industry gas supply shocks.} The x-axis shows the possible range of aggregate gas shocks. The y-axis is the gas supply shock in \% of pre-shock industry gas usage. We find ranges where no aggregate shock is transmitted to the industry due to measures like savings from heating. Industries experience a 100\% shock already for aggregate shocks of 58\% and higher due to prioritization of other customers. 
    }
    \label{fig:gas-consume-type}
\end{figure}

Figure~\ref{fig:gas-consume-type}C shows how the country-wide gas supply shock translates into industry gas supply shocks when taking these assumptions into account. For country-wide gas supply shocks above 58\%, the remaining gas supply is less than the demand by protected consumers, and no gas is available for industrial consumers. For aggregate gas supply shocks below 12\%, the economy would remain unaffected as the reduction of the overall gas supply is compensated by the mitigation factors considered above. The slope of the curve for shocks between around 12\% and 58\% is 2.2. In this region, an additional percent aggregate gas supply shock translates into an additional 2.2\% industry gas supply shock, indicating the sensitivity of industry gas supply reduction with respect to overall reductions in the country’s gas supply. Our computations show that there is a 10.4\% shock to industry gas supply in the EU-wide cooperation Scenario A, whereas there is a 53.3\% shock to industry gas supply in the uncoordinated Scenario B (see Table~\ref{tab:scen}).

\begin{table}[!htbp]
\begin{tabular}{|l|c|c|}
\hline
& {\bf Scenario A} & {\bf Scenario B} \\ 
& \textit{ EU-cooperation } & \textit{ Uncoordinated }\\ 
\hline
{\bf Annual gas consumption (Austria 2021) }         & \multicolumn{2}{c|}{ {\bf 100\%} (9.34 bcm) } \\ 
\hline
- Russian import stop                            & \multicolumn{2}{c|}{-80\% (-7.47 bcm)} \\ 
\hline
+ Additional imports & \multicolumn{1}{l|}{55.7\% (5.20 bcm)}& 28.4\% (2.65 bcm)                                                  \\ \hline
+ Storage                                        & \multicolumn{1}{l|}{6.9\% (0.64 bcm)}                                                     & 15.0\% (1.40 bcm)                                                  \\ \hline
{\bf Country-wide gas supply shock }  & {\bf-17.4\%} (-1.63 bcm)                                                  & {\bf-36.6\%} (-3.42 bcm)  \\ 
\hline
+ Fuel switching         & \multicolumn{2}{c|}{10.5\% (0.98 bcm)} \\
\hline
+ Household savings in heating          & \multicolumn{2}{c|}{1.2\% (0.11 bcm)} \\ 
\hline
+ Industry savings in heating          & \multicolumn{2}{c|}{0.6\% (0.06 bcm)} \\ 
\hline
+ Savings from operating  pipelines              & \multicolumn{1}{l|}{0.5\% (0.05 bcm)} & 1.2\% (0.11 bcm) \\ 
\hline
{\bf Realized aggregate gas supply shock} & {\bf -5.2\% }(-0.49 bcm) & {\bf -23.7\%} (-2.21 bcm)  \\ 
\hline
{\bf Industry gas supply shock} & {\bf -10.4\%} & {\bf -53.3\% } \\ 
\hline
\end{tabular}
\caption{ {\bf Summary of gas supply shock scenarios.} Both scenarios assume a complete stop of gas imports from Russia as of June 1, 2022, translating into an annualized reduction of 7.47 bcm in gas imports. The two scenarios differ with respect to the amount of alternative imports secured, storage management and potential savings from pipeline operations. Percentages of \textit{Industry gas supply shock} refers to the relative reductions in industry gas supply. All other percentage values are expressed with respect to annual gas consumption (2021 numbers). Fuel switching relates to the electricity and CHP only. Details on the calculations and scenarios are found in Appendix~\ref{apx:mitigate}.
}
\label{tab:scen}
\end{table}

%%%%%%%%%%%%%%%%%%%%%%%%%%%%%%%%%%%%%%%%%%%%%%%%
\section{Direct economic shocks from gas shortages}
\label{sec:drect}

Figure~\ref{fig:gas-consume-type}A shows the gas consumption in Austria across consumer types according to the energy balance. Households and small consumers represent only about 19.5\% of total gas consumption. The industrial sectors contribute to approximately 44.1\% of annual gas consumption. When combined, heat and power (CHP) along with electricity account for an additional 26.3\%. The remaining 10.1\% is distributed among the service and public sector, pipeline transport, and heating plants. Fig.~\ref{fig:gas-consume-type}B zooms into industrial gas usage. Note that 28\% of industrial gas consumption is used for room heat. Gas shortages for room heat will likely have less severe consequences on economic production than shortages in process heat (accounting for 62\% of industry gas consumption) or non-energetic use, e.g., as a chemical reactant (10\%).

\subsection{Sectoral gas dependencies} \label{sec:gasdepend}
Using the material input surveys \citep{stat_guetereinsatz}, we estimate the gas dependency of economic production in each sector to quantify the direct economic consequences of the computed gas supply reductions in the industry. 
The data shows that \textit{Manufacture (M.) of basic iron and steel}, \textit{M. of paper and paperboard} and \textit{M. of organic chemical raw materials} are among the top consumers of natural gas in the country, whereas substantially less gas is consumed in other sectors. 
The amount of gas consumed, however, does not necessarily tell us about how dependent the economic production of a sector is in case of supply reductions. To quantify the output-gas dependency of a sector, we estimate the number of firms in a given sector that rely on natural gas in production by combining the material input survey with structural business statistics.

Let $g_i$ denote the share of firms in sector $i$ that report gas usage in production. To estimate $g_i$, we compare the number of gas-using firms with the total number of firms in the sector while taking reporting thresholds and firm size distributions into account (see Appendix~\ref{apx:gasdepend} for details).
In Fig.~\ref{fig:industry_dependency}A we show the estimated share of firms per sector plotted against the sectors' relative output shares. Note that we computed the relative output share of sector $i$ as $s_i = x_i / \sum_i x_i$, where $x_i$ represents sector $i$'s gross output.
The figure makes clear that the share of firms using gas inputs in production varies substantially across sectors.
Based on our estimates, more than 70\% of firms in \textit{M. basic metals} use gas in production. The share of firms using gas is also high for the sectors \textit{M. paper} and \textit{M. furniture}. For a number of sectors, the share of firms using gas is small. For example, for the largest sector in terms of output, \textit{Construction}, less than 10\% of firms rely on gas inputs.

\subsection{Direct output shocks} \label{sec:directshock}

We use the overall gas shock to industrial sectors and their estimated gas dependency to quantify direct output shocks. We only apply direct gas supply shocks to the gas-using firms of a sector, as estimated in the previous section. Thus, firms that are not using any gas in their production do not experience any direct decline in their production capacity, but they could be affected by indirect impacts. 
By assuming a simple linear relationship between sectoral output and gas inputs, the direct output reduction due to a reduction in gas supply is given by 
\begin{equation} \label{eq:direct}
    \Delta x_i^\text{d} = \alpha_i g_i  \epsilon^G_i x_i,
\end{equation}
where $\epsilon^G_i \in [-1,0]$ is the gas input reduction to sector $i$ and $\alpha_i \in [0,1]$ a parameter indicating how strongly a sector's output depends on the availability of natural gas. 
We distribute total gas shocks to production evenly across all sectors, $\epsilon^G_i = \epsilon^G$, but emphasize that different assumptions could be considered, e.g., to investigate the impact of different rationing schemes.
While not modeled explicitly here, in practice, we expect that the actual gas shock distribution would depend on price and substitution effects that affect the firms' willingness and ability to pay for the available gas inputs.
For the short-time horizon considered, our baseline assumption is that natural gas is a critical input for firms that use gas in production, with a one-to-one dependence between their output and gas inputs. Thus, we set $\alpha_i = 1 \; \forall i$.
Given the substantial uncertainty surrounding the key parameter $\alpha_i$, we quantify the importance of this assumption in our sensitivity analyses provided in Section~\ref{apx:sense}. 

\begin{figure}[!ht]
    \centering
        \includegraphics[width=\textwidth]{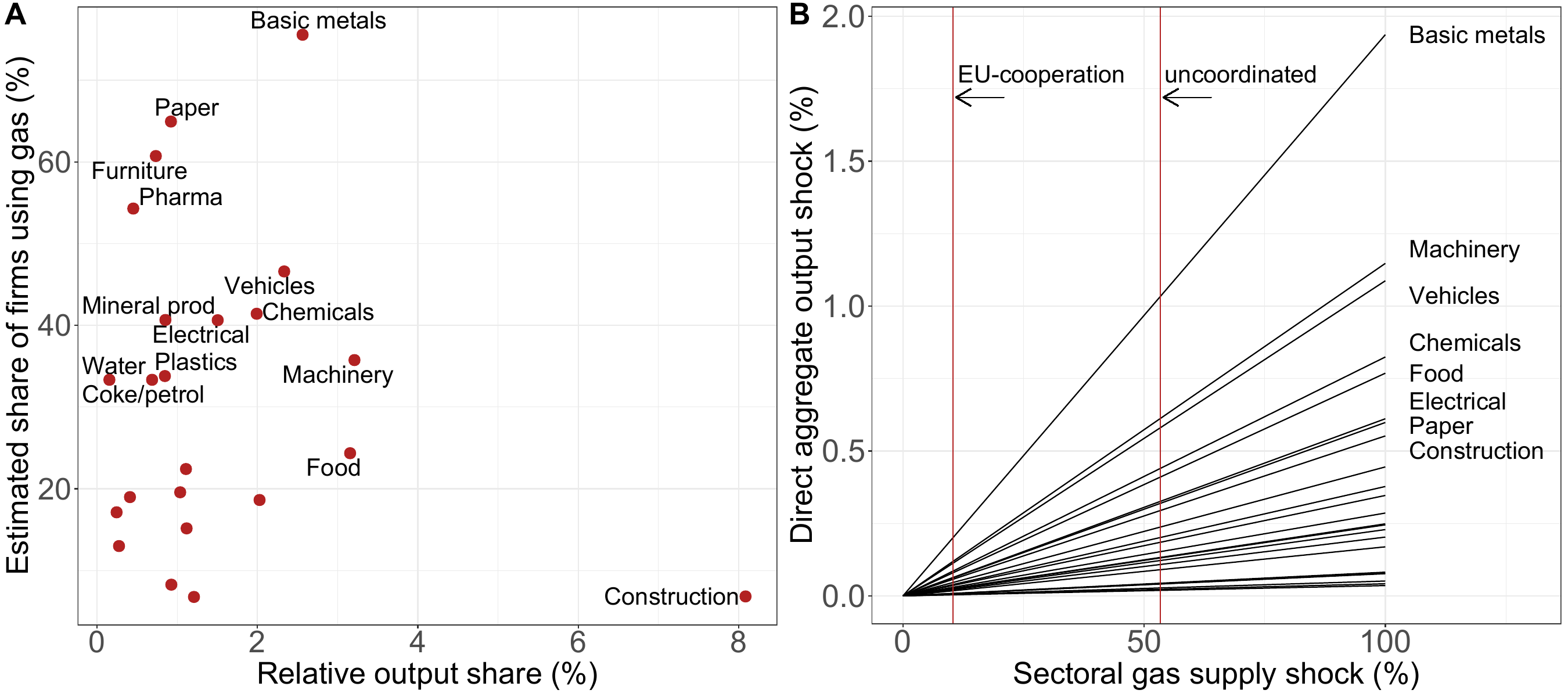}
    \caption{
    {\bf A: Estimated share of firms in a sector that uses gas $g_i$  versus sectoral relative output shares $s_i$.}
    Sector labels are custom abbreviations based on NACE classifications. All sectors labeled represent manufacturing industries, except for Construction.
    {\bf B: Direct sectoral output shocks induced by a gas supply shock $\epsilon^G_i$.}
    Note that the y-axis is expressed in terms of economy-wide output reduction, i.e., $\Delta x_i^\text{d} / \sum_i x_i$. We show labels for the eight most critical sectors in terms of direct gas shock impacts. The figure uses our baseline assumption $\alpha_i=1$.
    Note that we do not consider gas shortages to the sector \textit{Electricity, gas, steam, and air conditioning supply}, as gas supply to this sector is prioritized according to the national gas supply emergency order.
    }
    \label{fig:industry_dependency}
\end{figure}

We show direct output reductions expressed as $\Delta x_i^\text{d} / \sum_i x_i$ as a function of sector-specific gas supply reductions $G_i$ in Fig.~\ref{fig:industry_dependency}B. 
The figure indicates that \textit{M. basic metals} has the largest adverse direct impact on Austria's economy in case of gas input shortages, followed by \textit{M. machinery} and \textit{M. vehicles}. Despite a relatively low gas dependency of \textit{Construction}, the sector still exhibits significant potential for economy-wide output depression due to its large size. 
Note that the figure depends trivially on the parameter $\alpha_i$. For example, if we reduce the parameter by a factor of two, i.e., $\alpha_i = 0.5$, each curve's slope is also reduced by a factor of two.

The two vertical lines in Fig.~\ref{fig:industry_dependency}B indicate the two main scenarios considered. In the EU-cooperation Scenario A, where there are strong supply-side countermeasures put in place, the sum of all shocks to the industrial sector amounts to 1.1\% of Austrian gross output. In the uncoordinated Scenario B, however, industries face a substantially larger direct shock to output, amounting to 5.6\%.

%%%%%%%%%%%%%%%%%%%%%%%%%%%%%%%%%%%%%%%%%%%%%%%%
\section{Economic impacts of a drastic gas supply shock}
\label{sec:impact}

\subsection{Economic model}
To assess total economic impacts, i.e., direct shocks plus indirect impacts, resulting from a total import stop of Russian gas, we adapt a dynamic disequilibrium input-output model introduced by \cite{pichler2022forecasting}. 
This model was designed to quantify the upstream and downstream propagation of the sector-specific demand and supply shocks and builds on previous work on the response to natural disasters \citep{hallegatte2008adaptive, henriet2012firm, inoue2019firm}. 
In this model, sectors use input inventories for production to satisfy incoming demand from other sectors and final consumers. Sectoral production is limited through capacity and input constraints with inventory holdings serving as a buffer against output reductions due to shortages of intermediate inputs.
The model can be run with various sector-specific production functions and incorporates inventory dynamics, consumption, and labor market effects. 
The main outputs of the model are time series on a daily frequency of important macroeconomic variables such as gross output, profits, wages and salaries, and household consumption.
We adapt this model for gas shortages by explicitly considering industrial gas usage in production and initializing the model with data on the Austrian economy. In the model, the economy initially rests in a steady state until it experiences exogenous shocks, in this case, due to shortages in gas inputs. We give a detailed mathematical description of the model in Appendix~\ref{apx:model}.

We initialize the model with the shocks described in Section~\ref{sec:directshock}. The economic model is a short-term model where the adaptive capacity of economic agents is 
limited. Therefore, we simulate the economy over a relatively short time horizon of two months (on a daily resolution). As our baseline run, we use the partially binding Leontief (PBL) production function introduced in \cite{pichler2022forecasting}. 
To get lower and upper bounds of potential shock amplifications, we simulate the model with the Leontief and the linear production functions. 

In the Leontief case, each sector produces based on fixed production recipes, and a shortage in any of the intermediate inputs leads to an immediate reduction in output. This can result in a drastic downstream spread of shocks and, thus, provides an upper bound of adverse indirect economic impacts. 
The PBL function relaxes the Leontief assumption by positing that short-term production is only limited through the lack of \textit{critical} inputs, while it remains unaffected by the lack of \textit{non-critical} inputs.
In the linear production function case, all intermediate inputs are non-critical, and a sector continues production even if some inputs cannot be provided as long as there is a sufficient supply of the other inputs. Thus, the linear production function results in limited downstream shock propagation effects, providing a lower shock amplification bound. 

The model assumes that each sector aims to satisfy incoming demand even if it is not profitable in the short run. Orders to intermediate suppliers are placed based on fixed (empirically observed) input shares. As a consequence, upstream shock propagation mechanisms do not differ across the various production function specifications (see Appendix~\ref{apx:model} for details).

\subsection{Aggregate model predictions}
\begin{figure}[!ht]
    \centering
    \includegraphics[width=.6\textwidth]{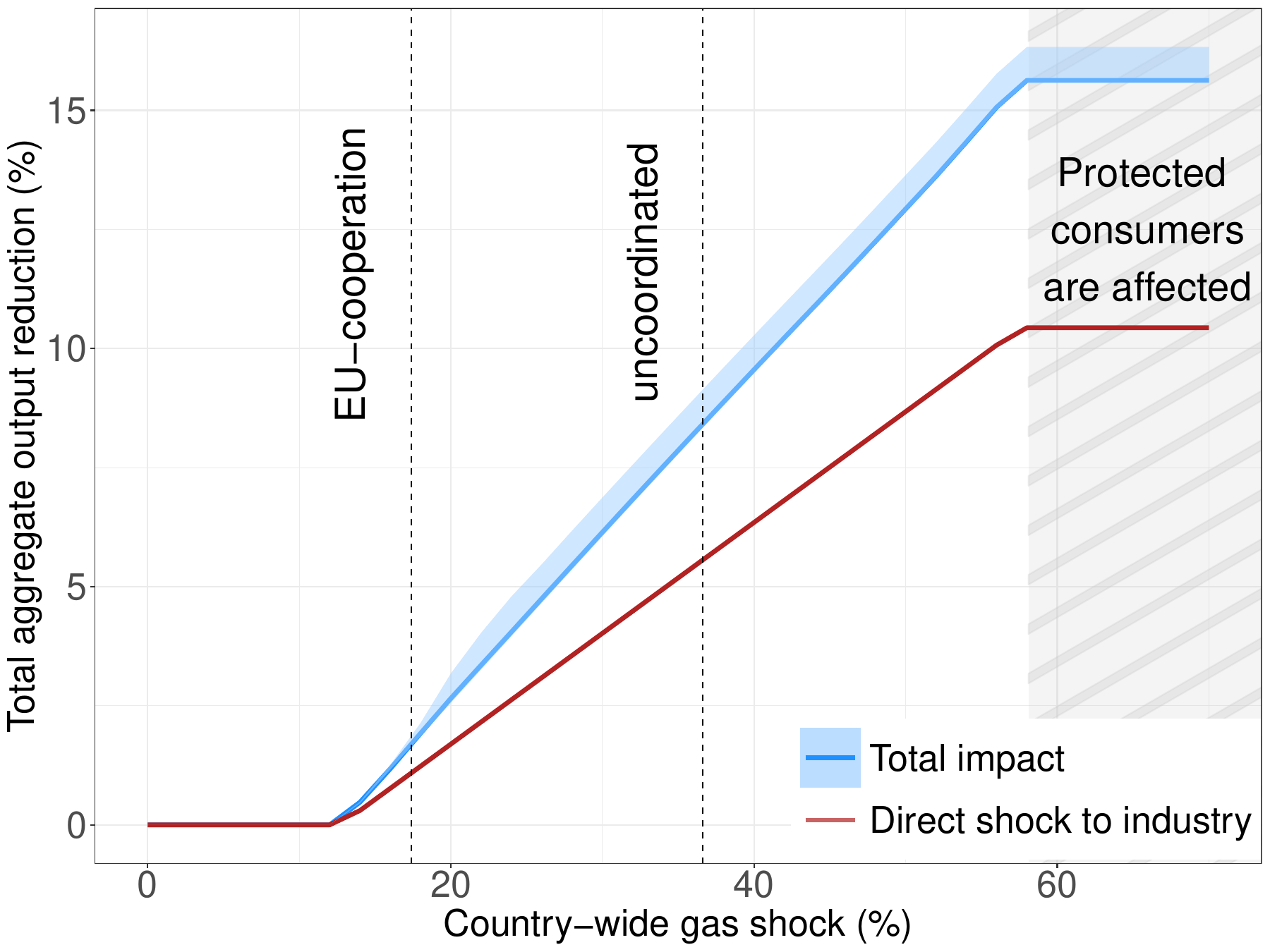}
    \caption{
    {\bf Aggregate output reduction due to country-wide gas supply shocks.}
    The y-axis shows aggregate output reductions, and the x-axis shows country-wide gas supply shocks. The red line shows aggregate direct economic shocks as computed in Section~\ref{sec:directshock}. The blue line and ribbon depict total output reductions after simulating the economic model for two months. The blue line corresponds to our baseline simulation using the PBL production function. The lower (upper) bound of the ribbon is obtained from running the model with a linear (Leontief) production function. 
    The two vertical dashed lines indicate the economic impacts of the main scenarios considered in this analysis. The grey area to the right corresponds to extreme gas supply shocks where also protected consumers, such as electricity, would experience gas supply reductions. We have not modeled the economic impacts of these drastic shock scenarios. 
    }
    \label{fig:key_econ_impact}
\end{figure}

We first focus on aggregate output reductions after two months due to reductions in natural gas, i.e., $1 - \frac{X_{t=60} - X_{t=0}}{X_{t=0}}$, where $X_{t} = \sum_i x_{i,t}$ denotes the total production of all sectors at day $t$ and $t=0$ indicates the pre-shock steady state.
Fig.~\ref{fig:key_econ_impact} shows the reduction in total gross output (y-axis) for the whole range of potential country-wide gas supply shocks (x-axis). The red line indicates aggregate direct shocks as computed in Section~\ref{sec:directshock}, and the blue line the total impacts after simulating the model with the PBL production function (the lower ribbon bound refers to linear, the upper bound to Leontief production).

The two dashed vertical lines in Fig.~\ref{fig:key_econ_impact} indicate the two main scenarios presented in Section~\ref{sec:gasscenario}. 
The left line, indicating the EU-cooperation Scenario A, represents a -17.4\% country-wide gas supply shock. This shock translates into a -1.1\% direct shock to output and an upper bound of -1.7\% total aggregate impact after indirect effects are considered. For the uncoordinated Scenario B, representing a 36.6\% country-wide gas supply shock, we observe a -5.6\% direct output shock and a -8.4\% overall negative impact on gross output. These results emphasize the importance of supply-side measures in mitigating the adverse economic impacts of a drastic gas supply shock. In particular, for a small, open, and landlocked economy like Austria exhibiting disproportionately large gas dependencies to Russia, the ability to source natural gas from alternative trade partners and storage management are key factors in curtailing adverse economic impacts.

For small aggregate gas shocks $< $14\%, our analysis suggests no adverse economic impacts as these supply reductions could be mitigated by demand-side savings without constraining economic production (see Section~\ref{sec:gasscenario}). For a country-wide gas supply shock $\ge 58\%$, no gas is available for industrial production as non-industrial consumers (households, public sector and energy) are prioritized according to the national gas emergency order. We do not model gas supply shocks to prioritized consumers explicitly since our analysis in Section~\ref{sec:gasscenario} suggests that such extreme shock scenarios are highly implausible. Thus, our simulations result in maximal adverse impact at this threshold (-10\% direct shocks, -16\% aggregate impact). However, note that impacts would further amplify beyond this threshold, in particular, if gas shocks to the energy sector, a critical input to most other sectors, would be considered. As demonstrated by \cite{leahy2012cost} in the case of severe gas supply disruptions in Ireland, shock amplifications can be enormous if electricity-generating gas plants are affected.

Figure~\ref{fig:key_econ_impact} shows that aggregate gross output is highly sensitive with respect to country-wide gas supply shocks falling into the range between 14\% and 58\%, which represent drastic but plausible shock scenarios given Austria's high dependency on Russian gas.
In this range, an additional 1\% decrease in aggregate gas supply translates into a $\sim$0.25\% decrease in gross output, indicating that even small improvements in gas availability can have a substantial positive economic impact. 
Similarly, we can evaluate the marginal economic benefit resulting from reducing the room temperature in residential homes. In the case of the severe gas supply disruptions considered here, a reduction in the average room temperature of 1$^{\circ}$C results in about 0.3\% additional output, also highlighting the welfare tradeoff involved in prioritizing different consumer types.
These results further emphasize that policies aimed at enhancing gas supply can have significant economic benefits, even if they are costly. 

In stark contrast to other studies, our results are highly robust with respect to the choice of the production function (as indicated in Fig.~\ref{fig:key_econ_impact} by the relatively narrow blue ribbon). The economic model amplifies direct shocks by a factor of about 1.5 when assuming linear production function compared to an amplification of about 1.6--1.8 under Leontief production. This observation suggests that there is limited downstream shock amplification in our simulations, even when using Leontief production functions for every sector. Thus, most shock amplification observed is substantially caused by upstream effects (reduced demand to suppliers) and Keynesian income effects. We further discuss shock propagation dynamics in Section~\ref{sec:shockprop} and investigate the effects of different shock amplification channels in more depth in sensitivity analyses (Appendix~\ref{apx:sense}).

We show model predictions of other aggregate variables for the EU-cooperation Scenario A and the Uncoordinated Scenario B in Table~\ref{tab:restab}. The table shows that impacts vary substantially across different economic variables.
In both scenarios, the model predicts the largest drops in aggregate profits, amounting to more than -10\% in the more severe gas supply shock Scenario B. Aggregate wages and salaries, which in our model are equivalent to total employment, experience the smallest impacts, resulting in about a 0.5\% and a 2.2\% reduction in Scenario A and Scenario B, respectively.

\begin{table}[ht]
\centering
\begin{tabular}{|l|rrr|rrr|}
  \hline
  & \multicolumn{3}{c|}{\textbf{EU-Cooperation}} & \multicolumn{3}{c|}{\textbf{Uncoordinated}} \\
 & Leontief & PBL & Linear & Leontief & PBL & Linear \\ 
  \hline
Gross output & -1.9 & -1.7 & -1.6 & -9.1 & -8.4 & -8.3 \\ 
  Profits & -2.8 & -2.6 & -2.6 & -12.1 & -11.1 & -11.1 \\ 
  Wages and Salaries & -0.6 & -0.5 & -0.4 & -2.9 & -2.2 & -2.1 \\ 
  Private consumption & -1.1 & -0.7 & -0.7 & -4.8 & -3.3 & -3.1 \\ 
   \hline
\end{tabular}
\caption{{\bf Model predictions of key aggregate variables for the two main scenarios}. 
Percentages refer to changes after two months of model simulations compared with the pre-shock economic situation. Results are shown for three different choices of production functions.
}
\label{tab:restab}
\end{table}

\subsection{Sectoral model predictions} \label{sec:sectorresult}
\begin{figure}[!ht]
    \centering
    \includegraphics[width=\textwidth]{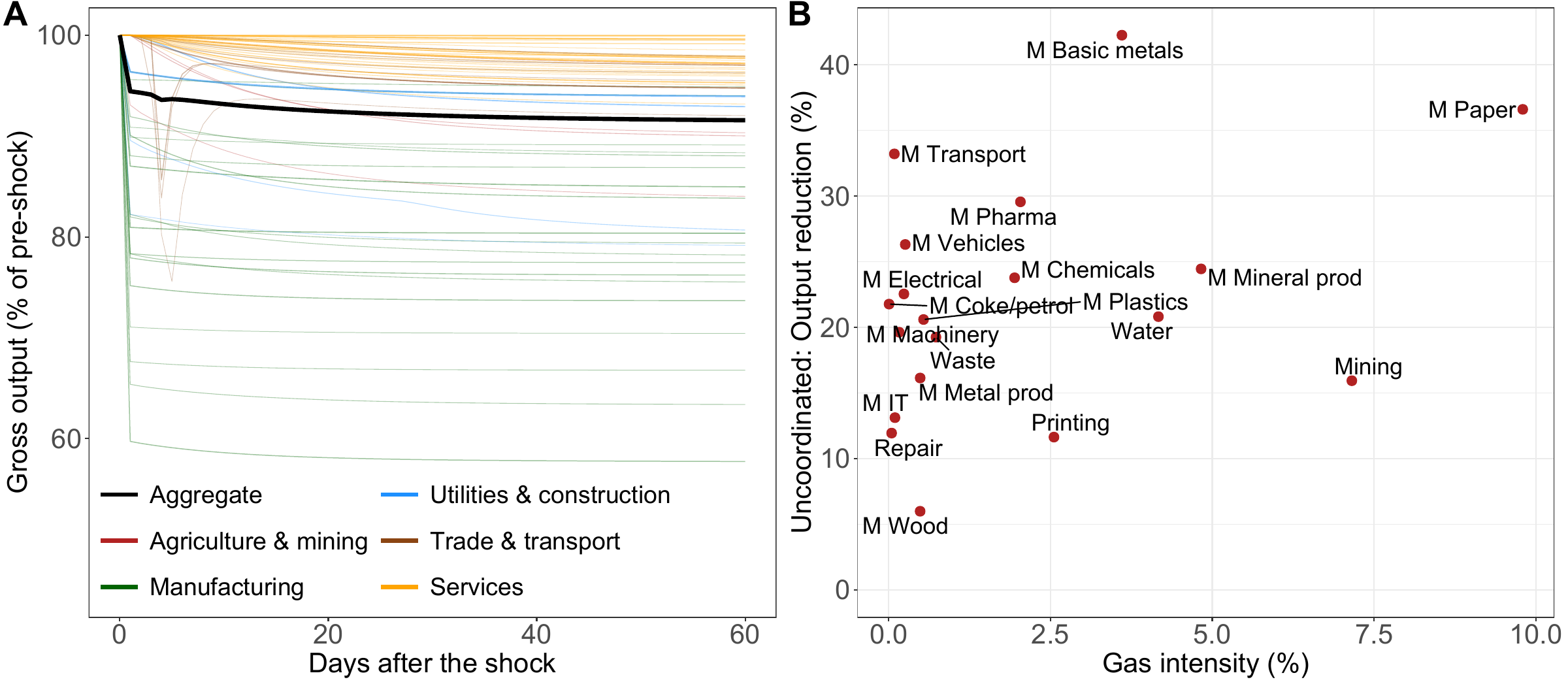}
    \caption{
    {\bf A: Production dynamics.}
    The figure shows sectoral and aggregate output in \% of pre-shock levels for the Uncoordinated Scenario B. Line widths are scaled to initial sectoral output levels.
    {\bf B: Estimated sectoral output reductions vs. gas intensity.}
    The y-axis shows output reductions for the Uncoordinated Scenario B two months after the gas shock. The x-axis shows gas intensity computed as sectoral gas consumption in cubic meters per sectoral output in EUR. 
    Only sectors that report gas usage according to the material inputs surveys are shown. Note that gas intensity for some sectors has been estimated (see Appendix~\ref{apx:gasdepend}).
    }
    \label{fig:sect_econ_impact}
\end{figure}

We show the dynamics of sectoral and aggregate production for the Uncoordinated Scenario B in Fig.~\ref{fig:sect_econ_impact}A. While several manufacturing sectors suffer substantial production losses, most service sectors experience only minor reductions in output. 
Sectors hit hardest, such as \emph{M. basic metals}, exhibit minimal production adjustments throughout the simulation, as they are primarily constrained by gas inputs. 
In contrast, several other sectors, particularly sectors in services and utilities, experience further production declines following the gas supply disruption due to reduced demand.
As noted, only a few sectors, all part of transport, encounter input bottlenecks in our simulations. Input bottlenecks arise soon after the gas supply shock occurs, resulting in sharp output declines followed by rapid recoveries. Despite substantial sectoral heterogeneity, aggregate production declines relatively smoothly and slowly over the two months after the gas import shock.

Figure~\ref{fig:sect_econ_impact}B shows that we predict the highest output reductions in \emph{M. basic metals}, followed by \emph{M. paper and paper products}, \emph{M. of transport equipment}, \emph{M. of pharmaceutical products}, \emph{M. of vehicles} and \emph{M. of non-metallic mineral products}. 
Unsurprisingly, the industrial sectors most adversely affected are those with the highest proportion of firms relying on gas in production, as shown in Fig.~\ref{fig:industry_dependency}A. This observation does not imply that gas-intensive sectors are necessarily the hardest hit. As shown in Fig.~\ref{fig:sect_econ_impact}B, there is no statistically significant linear relationship between output reductions and gas intensities. This is because, in our analysis, gas shortages are evenly distributed across all sectors, and firm outputs are assumed to be equally sensitive to gas shortages, regardless of gas intensities.

\FloatBarrier
%%%%%%%%%%%%%%%%%%%%%%%%%%%%%%%%%%%%%%%%%%%%%%%%
\section{Discussion}
\label{sec:discuss}

\subsection{Shock propagation dynamics} \label{sec:shockprop}

Our results are remarkably robust to the choice of the production function since we observe limited downstream propagation (upstream propagation is the same across production function choices). Even with Leontief production, where shortages of any input reduces production, input bottlenecks are rare. The scarcity of bottlenecks can be explained by the following factors: the presence of inventories, the limited scope and magnitude of shocks, and the predominance of other constraints.

In our model, industries maintain input inventories, allowing them to continue production in the short term if their suppliers fail to deliver sufficient inputs. These inventories act as buffers against shocks arising upstream in the supply chain. Downstream propagation only occurs if inventories fall at low levels that are insufficient to meet incoming demand. 
In the model, sectors reduce their demand for inputs and reduce their productive capacity through layoffs when they encounter reductions in demand or input constraints. These adjustments propagate shocks upstream, further reducing the likelihood of input bottlenecks.
In our baseline simulations, we observe that three transport sectors face input bottlenecks, while production in other sectors is constrained by either gas inputs or demand. Thus, other constraints tend to dominate input constraints. This is not obvious. Simulating the economic impacts of the COVID-19 pandemic \citep{pichler2022forecasting} with a similar model results in substantially higher downstream shock propagation. Compared to the pandemic shocks, however, the gas shocks considered here are relatively small and localized.

We have initialized the sectoral inventory holdings using U.K. data from \cite{pichler2022forecasting}, since we could not access comparable data for Austria. In this data, average inventory levels are about 25 days of production.
Relying on foreign inventory data entails uncertainties around the inventory parametrization in our model. To better understand the impact of inventory levels on model results, we have conducted sensitivity analyses detailed in Appendix~\ref{apx:sense}. We find that the results presented here are quite robust even with substantially lower inventory levels. Halving the baseline inventory holdings, the model predicts up to 1\% lower output than our baseline simulations (and up to 3.5\% with the more stringent Leontief production function).
We observe substantial heterogeneity across sectoral inventory holdings, ranging from less than 2 to over 70 days of production. As demonstrated in Appendix~\ref{apx:sense}, the heterogeneity of inventory levels increases shock propagation.

One model assumption that limits downstream shock propagation is the production of homogeneous goods within sectors. Although shocks are applied only to the sector share that uses gas, we assume that firms within a sector produce the same type of good, regardless of gas usage. However, firm heterogeneity within a sector can significantly amplify shocks \citep{diem2024estimating}. For example, \cite{hallegatte2014modeling} introduces a model where sectoral production is reduced as soon as inventories drop below the target inventory level, with an additional parameter controlling intra-sectoral firm heterogeneity. Greater intra-sectoral heterogeneity results in larger shock amplification. While we did not explore how alternative modeling choices affect our results, we consider a systematic comparison of various downstream propagation mechanisms to be an interesting future research avenue.

\subsection{Limitations of economic impact analysis} \label{sec:limits}

A major limitation in our analysis is that our economic model assumes time-invariant prices, no shifts in consumer preferences, and no changes in international trade. Since 2021, we have observed substantial increases in European gas prices, incentivizing firms to reduce demand, substitute gas-intensive products with imports, and invest in alternative energy sources. After controlling for temperature effects, \cite{ruhnau2023natural} estimate that German gas consumption dropped by 23\% in the second half of 2022, although it is impossible to disentangle between price effects and crisis effects such as increased public attention due to the Russian invasion of Ukraine on February 24, 2022. Note that we have incorporated gas consumption savings unrelated to economic production exogenously in the development of our gas shock scenarios (Section~\ref{sec:gasscenario}), which, however, have been rather conservative (households reducing demand by about 6\%).

Given the lack of additional data, we have assumed no possibility of gas substitution in the industry for the short time horizon considered. An exception is the energy sector, where we could make informed estimates on the potential of fuel-switching for specific plants due to their technical characteristics. We argue that this assumption is reasonable for many industrial applications where substituting gas with other inputs would require replacing or retrofitting existing capital stock, which is likely extremely difficult and costly in the short term. Nevertheless, we acknowledge that several industrial processes might exhibit a larger potential for gas substitution, which we could not consider. For example, \cite{stiewe2022european} show that German fertilizer production, which is highly dependent on natural gas to produce ammonia, responded with higher imports of ammonia to higher gas prices.
\cite{moll2023power} provide a number of further empirical examples showcasing how German firms have adapted to increased gas prices.

Due to limited data availability, we had to estimate the gas dependency for each sector based on firm counts, which is obviously imperfect. Similarly, without further details on underlying production functions, we assumed a simple linear relation between productive capacity and gas inputs for gas-using firms, even though, in reality, we expect substantial heterogeneity across firms. Since our economic model operates at the sector level, we assume that each sector produces a homogeneous good, potentially underestimating the heterogeneity of shock propagation. Thus, our model would not allow us to detect cases where gas-using firms produce very specific goods that represent critical inputs to other firms. While this limitation applies to most macroeconomic models, in principle, our model could be calibrated to firm-level production network data, if such data were available.

\subsection{Comparison with empirical gas imports and consumption} \label{sec:empirical}
Our computations have been based on a full export ban of Russian gas to Europe which did not materialize, thus, representing counterfactual scenarios. It is nevertheless informative to revisit our computations in light of the empirical evolution of gas usage, to better understand the plausibility of various assumptions. Since our results are based on a prospective study of severe gas supply disruptions in June 2022, we contextualize them here by discussing the observed changes in gas usage between 2021 and 2022 (see Table~\ref{tab:comparison}).

While we assumed a complete stop of Russian gas imports to the EU (155 bcm in 2021), actual gas imports dropped to 74 bcm in 2022 and further declined throughout the year (between June 2021 and June 2022 Russian gas imports even dropped by 124 bcm). 
These reductions were not as sudden as assumed in our calculations and were not uniform across transit routes.
Gas transit through the Yamal pipeline, passing through Poland and Germany, and the North Stream pipeline, entering Germany from the sea, completely stopped in the middle of 2022. 
In contrast, gas transit through the Turkstream pipeline (passing through Bulgaria) only decreased by 7\%, and gas transit through Ukraine, which represents the main route for Russian gas to Austria, decreased by 67\% (SI Section~S2).

Assessing how Austria's gas imports from Russia have changed is more difficult since, to the best of our knowledge, no consistent time series of Russian gas imports exists that comprehensively covers the time before and after 2022.
Using monthly data that has become recently available \citep{bmk_ru}, we show in the SI Section~S1 that Austria's total gas imports have significantly declined during 2022. For some months in mid-2022, the Russian gas import share dropped to levels around 20\% but experienced substantial increases afterward. Between September 2023 and April 2024, Austria's monthly gas imports consisted of at least 76\% of Russian sources and reached levels up to 97\%. Thus, any observed reductions in Russian gas import shares so far have only been temporary. 
while the import share of Russia is still extremely high (up to 97\%).

We have built our scenarios upon the RePowerEU plan \citep{eu22_repower}, which provided us with EU-level projections on additional gas imports from non-Russian sources (Appendix~\ref{apx:_mit_imp}). While the EU Commission targeted additional imports of 50 bcm of LNG and 10 bcm of gas from non-Russian pipelines, the actual numbers were 55 bcm of LNG and 30 bcm of gas from pipelines, exceeding prior expectations \citep{fulwood2022eu}.

In our analysis, we assumed that heating-related behavioral changes could reduce Austria's overall gas consumption by 1.8\%. While we lack numbers for Austria, it has been estimated that behavioral changes and fuel switching led to a 2\% reduction in gas demand in the building sector in the EU \citep{zeniewski2023europe}. Since our scenarios have been developed in the context of a severe gas supply disruption (that did not materialize), our estimates of heating-related gas-saving potential have been rather conservative. We have intentionally adopted a conservative perspective because heating-related gas consumption is heavily influenced by climate factors such as winter temperatures, and a cold winter could easily offset savings from behavioral changes in heating. In the EU, 5.3\% of observed gas demand reductions can be attributed to the warm winter 2022-2023 \citep{zeniewski2023europe}. 

\cite{zeniewski2023europe} further estimate that around 1.7\% of gas savings in the EU were due to increased coal burning in the power sector, compared to our assumption of 10.5\% gas-saving potential due to fuel-switching. Since the gas supply disruption considered in our analysis did not occur, the need for fuel switching was less pronounced. Consequently, we would not expect our scenario assumptions to align with the observed data.

To summarize, our computations rest on much more extreme gas supply disruptions than actually witnessed during the recent energy crisis. Although there have been significant reductions in gas imports from Russia to the EU, these occurred gradually over extended periods. While the EU has made considerable progress in diversifying its gas import sources in line with our scenario assumptions, countries like Austria remain to heavily rely on Russian natural gas. In the event of a sudden and more severe gas supply disruption, we anticipate that the potential for heating-related savings could be greater than what we have assumed. However, highly contingent driving forces such as winter temperatures necessitate a conservative estimate when planning gas supplies ahead of time.

\begin{table}[!h]
    \centering
    \resizebox{\textwidth}{!}{  
\begin{tabular}{|l|c|c|c|l|}
\hline
\textbf{Changes   in …}                               & \textbf{Geo} & \textbf{Scenarios} & \textbf{Empirical} & \textbf{Source} \\ \hline
Imports of   Russian gas                              & EU           & -155 bcm            & -81 bcm           &      \cite{bruegel_data}   \\ \hline 
LNG Imports                                           & EU           & +50 bcm             & +56 bcm           &      \cite{bruegel_data}        \\ \hline
Imports from   non-Russian pipelines                  & EU           & +10 bcm             & +30 bcm            &    \cite{bruegel_data}         \\ \hline
Gas   consumption due to behavioral changes           &              & -1.8\% (AT)         & -1.7\% (EU)        &      \cite{zeniewski2023europe}        \\ \hline
Gas   consumption due to fuel-switching and more coal &              & -10.5\% (AT)        & -1.7\% (EU)        &     \cite{zeniewski2023europe}         \\ \hline
\end{tabular}
}
    \caption{ {\bf Comparison of scenario assumptions and actual empirical changes in gas imports and consumption.}
    Details on scenario values are found in Appendix~\ref{apx:mitigate}. The column \emph{Empirical} reports observed differences between annual values from 2022 with 2021. Note that the comparison is only indicative as our analysis has been conducted for gas supply disruptions occurring in mid-2022. Additionally, due to data limitations, we had to compare EU data with Austrian data for some variables.
    }
    \label{tab:comparison}
\end{table}

\FloatBarrier
%%%%%%%%%%%%%%%%%%%%%%%%%%%%%%%%%%%%%%%%%%%%%%%%
\section{Conclusion}
\label{sec:conclude}
Here, we analyzed the possible economic impacts of a sudden stop of Russian gas exports to Austria as of June 1, 2022. We first estimate the additional availability of natural gas for Austria in two illustrative scenarios, an EU-cooperation Scenario A and an uncoordinated Scenario B, where Austria tries to secure additional gas individually. We then quantify the likely reduction of natural gas available for industrial production and simulate the overall economic impact on the Austrian economy with a dynamic out-of-equilibrium production network model. 

Results show that through coordinated policy measures on the supply and demand side, adverse economic consequences of a drastic gas shock can be substantially reduced. The reduction in production in the EU-cooperation scenario is about a factor of five smaller than in an uncoordinated scenario. We observe a high sensitivity of gross output to a plausible range of gas shocks given the threat of a sudden stop in Russian imports in mid-2022. In this range, an additional 1\% decrease in aggregate gas supply translates into a 0.25\% (about 150 million EUR per month) decrease in gross output. These results indicate that in case of a drastic gas supply shock, policies resulting in modest improvements in gas availability can have substantial positive effects on the economy, even when these measures are costly.

Our analysis identifies several effective policies to mitigate adverse economic impacts from a sudden, drastic gas supply shock. Securing gas supplies from alternative trade partners has been among the key priorities of policymakers in preparing against potential gas import disruptions. 
We have studied two scenarios representing different EU-level policy decisions, resulting in very different additional gas supplies for small and disproportionately exposed countries like Austria.
Our results confirm that securing alternative gas supplies could be a key policy variable to mitigate the short-term negative consequences of a Russian gas import stop. Since Austria does not have access to LNG ports and about 90\% of its annual gas consumption is imported via European countries, Austria disproportionately benefits from accessing the European gas market and an EU-wide coordination in gas supply policies.

Another key factor to mitigate short-term impacts is the use of gas storage. The Austrian government reacted soon after the Russian invasion of Ukraine on February 24, 2022 by issuing an amendment to secure ``strategic gas reserves'' of 20 TWh ($\sim$20\% of consumption in 2021), which can be provided in times of crisis. Moreover, the government invested in securing additional gas reserves for protected consumers, and a substantial share of Austria's gas storage facilities are secured as ``immunized reserves'', i.e., reserves owned by large industrial consumers that they can access if the gas emergency order (``Energielenkungsgesetz'') becomes effective. Nevertheless, as of early 2024, almost half of Austria's gas reserves are held by foreign owners. Thus, in case of Europe-wide gas supply shortages, these reserves might not be available for domestic consumption, further emphasizing the importance of a coordinated crisis response at the level of the European Union.

Our analysis also demonstrated the large potential of substituting gas with alternative fuels. In our crisis scenario development, we have assumed fuel substitutions in gas power plants amounting to annual gas consumption savings of about 1 bcm. Similarly, the reactivation of already phased-out Coal plants, such as mandated by the Austrian government in the summer of 2022, can lead to substantial substitutions of gas inputs in the electricity sector within several months. Due to the lack of more granular data, we did not consider fuel substitution in the industrial sectors. However, as evidenced by significant reductions in industrial gas consumption and several anecdotal examples, there is likely substantial potential for gas substitution for a number of industrial processes. Note that less than 1,400 companies in Austria are required to report gas usage in the material input surveys. Thus, it seems feasible for a policymaker to understand industrial gas substitution potentials based on first principles by engaging directly with the stakeholders and by collecting more detailed data on gas usage. 
Such data would reduce the need to resort to aggregate macroeconomic models based on abstract ad-hoc assumptions on substitution and contribute to more realistic assessments of economic impacts at the macro- and micro-level.

Another policy lever is to encourage gas saving in activities unrelated to economic production, for example, through public campaigning (which is part of the ``Early warning level'' of the Austrian Gas Emergency Plan). A successful example is Japan after the Fukushima nuclear accident in 2011, where large public campaigns contributed to significant and lasting savings in electricity consumption \citep{kimura2016responding}. As our analysis represented an ex-ante evaluation of a potential Russian gas export stop to Austria on June 1, 2022, we have made conservative assumptions on the saving potentials through consumption reduction in heating. While high energy prices have clearly contributed to further reductions in gas consumption \citep{ruhnau2023natural}, it is still important to note that heating-related gas savings will strongly depend on climate factors (such as winter temperatures) and, thus, are difficult to anticipate beforehand.

In the mid to long term, there is substantial potential to reduce the country's gas dependency through incentivizing the adoption of new technologies and improving energy efficiency. As outlined in \cite{aea2022strategische}, among the key factors to reduce Austria's dependency over the next years are the replacement of gas-based heating in households and industry with renewable heating systems (e.g., heat pumps), accelerated renovation of buildings and switching to renewable energy in industrial production as well as in the heating and power sector. Policy measures targeted at the diffusion of renewable energy sources and improved energy efficiency would not only contribute to preventing future energy crises but are also critical for achieving the national carbon neutrality target in 2040 \citep{iea2020austria}.

\bibliographystyle{agsm}
\bibliography{gasbib}

\appendix
\input{appendix/mitigation_strat}

\input{appendix/gas_dependency} 
\input{appendix/model}

\input{appendix/add_results}

% After compiling the SI as pdf this can be executed
%\FloatBarrier
%\includepdf[pages=-,pagecommand={},width=\textwidth]{SI/SI.pdf}

\end{document}

\begin{center}
\huge{\textbf{Supplementary Information}}
\end{center}
\vspace{0.05cm}

\begin{center}
\LARGE{Economic impacts of a drastic gas supply shock and short-term mitigation strategies}
\end{center}
\vspace{0.05cm}

\begin{center}
\large Anton Pichler, Jan Hurt, Tobias Reisch, Johannes Stangl, Stefan Thurner
\end{center}

\tableofcontents
\newpage

%%%%%%%%%%%%%%%%%%%%%%%%%%%%%%%%%%%%%%%%%%%%%%%%

\section{Gas supply and consumption}
\label{sec:gas_at}

\subsection{Consumption}
Since our analysis concerns a drastic gas supply shock in June 2022, we treat the gas consumption profiles of 2021, whenever data availability allows, as the baseline. One important feature of Austria's gas consumption is its disproportionately high dependency on Russian imports. 
To put the Austrian situation in perspective, note that Russian imports accounted for 38\% (155 billion cubic meters [bcm]) of the EU's annual gas consumption in 2021 while representing about 80\% (9.34 bcm) of Austria's yearly gas consumption \citep{stat_gesamtenergie} (see Section~\ref{si:at_import} for details). 

Overall, Austria's gas consumption amounted to 9.34 bcm in 2021 (5\% increase to 2020), representing 2.3\% of EU-wide consumption (Fig.~\ref{fig:gas-consume}A). 
Between 2016 and 2021 Austria’s annual gas consumption has remained relatively stable between 8.7 and 9.5 bcm (Fig.~\ref{fig:gas-consume}B). A large share of natural gas is used for heating and power generation, in particular in the winter months, leading to a strong seasonal component in intra-annual gas consumption (Fig.~\ref{fig:gas-consume}C). This seasonality in gas consumption implies that the timing of a possible gas supply shock is crucial for assessing its impact on the country's energy security. As shown in Fig.~\ref{fig:gas-consume}C, gas consumption has repeatedly dropped below historical minimum levels since mid-2022.

\begin{figure}[!h]
    \centering
    \includegraphics[width=\textwidth]{fig/gas_cons1.pdf}
    \caption{ {\bf A: EU-27 gas consumption in 2021 per member state.} 
     Austria only accounts for slightly above 2\% of EU’s consumption. 
     {\bf B: Annual gas consumption between 2014 and 2024.}
     {\bf C: Monthly gas consumption.} There is a strong seasonality in gas consumption. The gray shade indicates the monthly min-max range between 2014--2021 and the lines show monthly consumption levels for the years 2021--2024. All three panels are based on data from Eurostat (nrg\_cb\_gasm) and consumption data refers to ``Inland consumption - observed''.
    }
    \label{fig:gas-consume}
\end{figure}

\subsection{Supply} \label{si:at_import}
A key number for deriving the main scenarios presented in the main text is the Austrian dependency on Russian gas, i.e., the share of Russian gas in Austria's annual consumption. It has to be noted that there has been substantial uncertainty around this number when our prospective analysis was conducted in mid-2022 since different data sources could only provide partial insights into serving this number.

Only recently, the Austrian government started publishing the share of Russian sources in its gas imports (this data was not available in 2022). In Fig.~\ref{fig:gas-prod}A we show the monthly gas imports to Austria decomposed into Russian and other sources. The figure makes clear that Austria did not experience a sudden Russian gas import disruption in 2022. While gas imports to Austria declined substantially during that year, Russia still accounted for the vast majority of gas imports in 2023 and the early months of 2024 (up to 97\%). 

However, note that not all imports are directly consumed domestically as Austria re-exports substantial amounts of natural gas (Fig.~\ref{fig:gas-prod}B). Thus, we have estimated the share of Russian gas in Austria's consumption using data from the European Union Agency for the Cooperation of Energy Regulators (ACER). The ACER dataset ``Estimated number and diversity of supply sources'' lists the top three gas supply sources per year which can also include own extraction. In this data, we observe that for every reported year, Russia represents the main direct source of Austrian gas supply. The second largest origin is Germany (except for 2015 when it was Norway) and the third largest is domestic production. However, Germany also sources gas from Russia and, additionally, sources gas from the Netherlands which, in some years, also sources gas from Russia. Consequently, to estimate the Austrian gas dependency, we have to take the indirect exposures to Russian gas through other imports into account. 

Therefore, we approximate the Russian gas import share as
\begin{align}
%   & (1-\text{DP}) \times  \\ 
 \text{IS}_\text{RU} =  \\
    &\text{AT\_RU}  \\
    &
  + \text{AT\_DE} \times \text{DE\_RU} \\
  &
  + \text{AT\_DE} \times \text{DE\_NL} \times \text{NL\_RU} \\
  &
  + \text{AT\_DE} \times \text{DE\_NL} \times \text{NL\_DE} \times \text{DE\_RU},
\end{align} 
where $\text{AT\_RU}$ represents the direct share of Russian gas in Austrian gas supply, $\text{AT\_DE}$ the share of German gas in Austria, $\text{DE\_RU}$ the Russian share in German supply and analogous for the other variables. Again, using the ACER data, we compute Austria's gas dependency on Russia as
\begin{align}
    (1-\text{DP}) \times \text{IS}_\text{RU},
\end{align}
where $\text{DP}$ is the share of domestic production.

We show Austria's Russian gas import share and dependency in Fig.~\ref{fig:gas-prod}C, demonstrating a significant increase between 2016 and 2021. This data has been updated since the development of the scenarios in spring 2022. Our scenarios have been developed with earlier data which has shown a gas dependency of 80\% (based on the year 2019). To ensure consistency with the original policy brief, we have kept our initial computations, but emphasize that these estimates have changed ex-post. In particular, Austria's gas dependency on Russia increased to more than 90\% in 2021, suggesting that our assumptions of Austria's gas dependency were rather optimistic. More generally, as noted in the main text, the level of uncertainty in the data on gas production, usage, and trade is substantial, necessitating a careful interpretation of any analysis.

\begin{figure}[!h]
    \centering
    \includegraphics[width=.495\textwidth]{fig/RU_imports_edashboard.pdf}
    \includegraphics[width=.495\textwidth]{fig/AT_exp_imp.pdf}
    \includegraphics[width=.495\textwidth]{fig/RU_dependency_acer.pdf}
    \caption{ 
     {\bf A: Share of Russian imports since February 2022.}
The figure shows monthly gas imports to Austria decomposed into Russia (red) and other (blue) sources. The percentage share of Russian gas imports per month is displayed with white text within the red columns. Import shares are provided by the governmental ``Energiedashboard'' (which was not public at the time of developing the scenarios presented in the main text) and total import numbers have been computed based on Eurostat data (nrg\_cb\_gasm).
{\bf B: Austrian gas exports and imports.} Data source: Eurostat (nrg\_cb\_gasm).
{\bf C: The dependence of Austrian gas consumption on Russian supplies.} 
The red points indicate the import share of Russian gas in total Austrian gas imports. The blue triangles show the share of Russian gas in Austria's total annual gas supply, i.e., which we use as a proxy for gas dependency. For computations, see the main text. The figure is based on data from ACER.
    }
    \label{fig:gas-prod}
\end{figure}

\clearpage

\FloatBarrier
\section{EU gas import dynamics}
\label{apx: imports}
Our estimates, that the EU could import an additional +50 bcm of LNG in 2022 compared to 2021 were based on the EU's RePower goal.
Figure \ref{fig:eu lng imports} illustrates that LNG imports had already increased to the level required to achieve this goal prior to the outbreak of the war.
\begin{figure}[!htb]
    \centering
    \includegraphics[width=0.495\textwidth, trim={0cm 1.9cm 0cm .85cm},clip]{fig/eu27_lng_imports.pdf}
    \includegraphics[width=0.495\textwidth, trim={0cm 0cm 0cm 1.3cm},clip]{fig/eu27_gas_imports.pdf}
    \caption{{\bf Left panel: Weekly LNG imports of the European Union.} Already before the war the LNG-imports increased to the level necessary to reach the goal of the RePower plan to import an additional +50bcm compared to 2022.
    {\bf Right panel: Weekly pipeline and LNG imports of gas of the European Union.} The imports of Russian Gas (Russian pipelines are: North Stream, Yamal, Turkstream and Ukraine transit) decreased substantially through 2022.
    }
    \label{fig:eu lng imports}
\end{figure}

\newpage
%%%%%%%%%%%%%%%%%%%%%%%%%%%%%%%%%%%%%%%%%%%%%%%%
\section{Comparison with alternative approaches} \label{sec:compare}
In the aftermath of the Russian invasion of Ukraine on 24 February 2022, policymakers and scholars have widely debated the economic consequences of a sudden cut-off from Russian gas. The debate has been especially intense in Germany, which represented Europe's largest net importer of Russian gas before the start of the war, with estimates ranging from relatively mild economic impacts to the potential of the most severe recession since World War II (see, e.g., \cite{albrizio2022market}, \cite{bayer2022stopp}, \cite{krebs2022auswirkungen}). 

%The German situation, however, was quite different from the Austrian situation. While Austria exhibited (and still exhibits) much higher dependencies on Russian gas imports, it consumes by a factor of 10 less gas than Germany, only representing about 2\% of EU-wide gas consumption. Thus, the immediate gas shortage resulting from a sudden Russian import stop would have been disproportionately higher for Austria than for Germany. Nevertheless, due to its comparatively small size, substituting these shortages would be comparatively easier  -> whats the point?

Here, we provide some further context on how our analysis relates to these predictions by applying alternative economic modeling approaches. \cite{bachmann2022if} use a two-input constant elasticity of substitution (CES) production function to investigate how aggregate production is affected by the reduction of gas (or, more generally, energy) inputs. Following this approach, the decline in production $X$ due to a reduction in Gas inputs $\Delta G$ can be approximated with
\begin{equation} \label{eq:deltaX}
    \Delta \log (X)  \approx \alpha \; \Delta \log\left(G \right) \; + \; 0.5 \left(1-\sigma^{-1} \right) \alpha (1-\alpha)  \; \left( \Delta \log (G) \right)^2,
\end{equation}
where $\alpha$ represents the expenditure share of gas inputs and $\sigma>0$ denotes the elasticity of substitution between gas inputs and all other inputs. In the special case of zero-substitution between gas and any other inputs, $\sigma = 1$, the CES production function reduces to a Leontief production function, resulting in $ \Delta\log (X)  = \Delta \log (G)$.

Following this approach, we show how output $X$ is affected by reduced gas inputs under various assumptions of the elasticity of substitution $\sigma$ in Fig.~\ref{fig:bfmodel}A. The figure suggests that economic production highly depends on how easily natural gas can be substituted with other inputs. For high values of $\sigma$ around 0.2--0.3, impacts at aggregate outputs are largely confined even for extreme gas input shock cases.  
\cite{bachmann2022if} argue that $\sigma \approx 0.1$ is plausible, while $\sigma = 0.04$ represents a reasonable worst-case scenario. When using $\sigma = 0.1$, we obtain a change in gross output of -0.2\% for the EU-cooperation Scenario A and -3.3\% for the Uncoordinated Scenario B.
For the more conservative assumption of $\sigma = 0.04$, we obtain changes in gross output amounting to -0.3\% and -7.6\% for Scenario A and Scenario B, respectively. The figure does not show the limiting case $\sigma \rightarrow 0$, where the aggregate CES production function reduces to a Leontief production function. In this case, the estimated -10.4\% and -53.3\% gas supply shocks to economic production would translate into -10.4\% and -53.3\% output changes, respectively.

Building on \cite{baqaee2019networks}, \cite{bachmann2022if} propose a ``sufficient statistics'' to quantify the economic impacts of a drastic Russian gas import stop that has been used by a number of further studies (e.g., \cite{albrizio2022market}, \cite{berger2022potential}). According to this approach, the impact on gross national expenditure (GNE), i.e., private consumption + investments + government consumption, in a multi-sector general equilibrium model can be approximated by 
\begin{equation}
    \Delta \log (GNE)  \approx  \left( \frac{p_G m_G}{ GNE } \; + \; 0.5 \; \Delta \frac{p_G m_G}{ GNE } \right) \; \Delta \log (m_G), 
\end{equation}
where $m_G$ denotes gas imports and $p_G$ their price. While the gas expenditure share in GNE, $p_g m_G / GNE$ can be measured by data and $\Delta m_G$ be interpreted as the gas import shock, the key parameter in this model is given by $\Delta \frac{p_G m_G}{ GNE },$ the expected change in the GNE share of gas imports. 

In Fig.~\ref{fig:bfmodel}B, we show the change in GNE as a function of gas import shocks and various specifications of $\Delta \frac{p_G m_G}{ GNE }$. The figure illustrates that the model is quite insensitive with respect to gas import shocks as long as the change in the GNE share of gas imports is limited. 
We compute Austria's gas import share of GNE using annual GNE data provided by Statistik Austria, monthly average import prices provided by E-Control, and monthly gas import data provided by Eurostat. Our computations suggest that 
$\alpha = \frac{p_G m_G}{ GNE } = 1.35\%$ in 2020 but increases to $3.76\%$ in 2021 and to even $8.63\%$ in 2022. For 2023, we expect this share to decline substantially, but the 2023-GNE numbers are not yet available at the time of writing.

Note that these numbers vary substantially over time, even without experiencing a complete shutdown of imports from Russia (in fact, Russian sources accounted for between 76\% and 90\% of monthly gas imports to Austria between September 2023 and November 2023).
Using the 2020 value of gas import shares in GNE and assuming a quadrupling of Austria's gas import share due to a Russian gas import stop, the resulting change in GNE amounts to -0.4\% for the EU-cooperation scenario and -2.7\% for the Uncoordinated scenario. However, if we took the 2021 number of gas import shares in GNE and assumed a quadrupling of this number (which seems overly extreme given the high initial levels), it would result in -1.0\% for Scenario A and -7.2\% for Scenario B.

\begin{figure}[!h]
    \centering
    \includegraphics[width=\textwidth]{fig/B-F_two.pdf}
    \caption{    
    {\bf A: Dependency of production on gas inputs and the elasticity of substitution.} following \cite{bachmann2022if}.
    The x-axis represents the reduction in gas inputs, and the y-axis represents various levels of the elasticity of substitution $\sigma$ (between gas and all other inputs). Colors/contours represent changes in production. The figure's minimum level of $\sigma$ is 0.02 (note that this matters as results become sensitive for $\sigma \rightarrow 0$).      
    Note that $\alpha=0.014$, which is roughly equal to Austria's gas import share in GNE using 2020 data.
    {\bf B: Dependency of GNE on gas imports} following \cite{bachmann2022if}.
     The x-axis represents the reduction in gas imports, and the y-axis represents the change in the GNE gas import share $\Delta (p_g m_G / GNE)$ (between gas and all other inputs). Colors/contours represent changes in gross national expenditure (GNE). Note that the figure is shown for 2020 values of $p_g m_G / GNE \approx 0.014$.
    }
    \label{fig:bfmodel}
\end{figure}

\FloatBarrier
%%%%%%%%%%%%%%%%%%%%%%%%%%%%%%%%%%%%%%%%%%%%%%%%
%% Bibliography
\small
\bibliographystyle{agsm}
\bibliography{gasbib}

%% file: appendix/mitigation_strat.tex
\section{Short-term factors mitigating a drastic gas import shock}
\label{apx:mitigate}

\subsection{Alternative gas imports} \label{apx:_mit_imp}
To assess the potential of gas imports from alternative sources, we build upon the EU Commission's strategy to reduce Russian gas imports by 2/3 until the end of 2022 \citep{eu22_repower}. This strategy paper involves importing an additional 50 bcm of LNG during 2022, as well as importing an additional 10 bcm via existing pipeline infrastructure from Norway, Azerbaijan, and Algeria. The proposed strategy on tapping alternative import sources has been considered highly ambitious but feasible in principle \citep{fulwood2022eu}. The International Energy Agency (IEA) estimates a technical potential of an additional 60 bcm of LNG imports in the near term \citep{iea202210}. 
We take a moderately conservative approach and assume that the EU falls 10\% short of its goal of 50 bcm of additional imports of LNG by the end of 2022 while it realizes its additional 10 bcm pipeline import goal. Thus, additional EU-wide gas imports amount to 55 bcm, compensating for more than a third of Russian gas imports. 
While we take the European gas imports as given, we distinguish two cases of how much of the additional EU-wide gas supply can be accessed by Austria. 

\begin{itemize}
    \item In the EU-cooperation Scenario A, we assume that member states face a common shock and distribute existing and additional gas resources such that every country faces the same relative reduction in its gas supply. Given the overall reduction of 100 bcm and EU-wide gas consumption of 412 bcm, the gas supply reduction for every country amounts to 24.3\% compared to usual levels. The gas supply shock to Austria is then reduced from 80\% to 24.3\%, resulting in 55.7\% or 9.34 bcm * 55.7\% = 5.20 bcm of additional imports. 
    \item In the uncoordinated Scenario B, each member state individually tries to substitute its current Russian imports from other countries. This means that Austria would demand 7.47 bcm of gas from the market. Since only 55 bcm of 155 bcm of Russian gas can be compensated on an EU-wide level, we assume Austria would only receive 7.47 bcm * 55/155 = 2.65 bcm (28.4\% of annual demand). This corresponds to a scenario where each member state places demand equaling its Russian import shortfall on international markets. Due to constrained supplies, however, every member state is rationed on a pro-rata basis. 2.65 bcm of additional imports might be optimistic in an uncooperative scenario, as Austria depends strongly on the available capacities of pipeline and LNG port infrastructures of other countries, which might not be willing to pass through natural gas to foreign consumers. 
\end{itemize}

The difference in additional gas imports between the uncoordinated and the EU-cooperation scenario is substantial, amounting to about 27.3\% of Austria’s annual gas consumption. This emphasizes the high uncertainty associated with any estimate of economic impacts from a sudden stop in imports of Russian gas. Many different scenarios could arguably have materialized. Since the realized scenario depends largely on political factors that are impossible to predict, the extent to which Austrian gas availability will be reduced is not merely a question of market mechanisms and technical feasibility but also a key policy variable.

Timing plays a critical role in securing alternative imports. If the EU cannot increase its imports fast enough, for example, due to limited LNG terminal capacity, temporary shortages could exceed the annualized shortfall calculated in Table~\ref{tab:scen}. 
However, two factors mitigate this risk.
First, storage capacities, as discussed in Appendix~\ref{apx:_mit_store}, can buffer delays in ramping up alternative imports.
Second, as demonstrated in the SI, Section S2, the EU had already achieved the RePower target for LNG import rates well before the assumed gas shock.

\subsection{Storage management} \label{apx:_mit_store}
Austria’s gas storage accounts for almost 9\% of EU total storage capacities, and given its 2.3\% share of EU gas consumption, has a comparatively large capacity to smooth out varying gas in- and outflows. To compute the availability of gas from Austrian storage, we again distinguish between the EU-cooperation Scenario A and an uncoordinated Scenario B. In the former, we assume that the Austrian gas storage is only used for managing national consumption, while in the latter, we assume that the usage of gas storage is coordinated at a European level.

To understand the storage management potential for shock mitigation, we use a simple storage model. We assume that gas supply through imports and domestic production is constant throughout the year, i.e., the net inflow does not show any seasonality between the summer and winter months. Natural gas consumption, however, is strongly seasonal and higher in the winter than in the summer months. The difference in daily in- and outflow corresponds to the storage injection or extraction, respectively, and allows us to predict daily storage levels. To keep a safety buffer even in the case of a sudden import stop from Russian gas, the model assumes extraction would be managed such that the storage level is expected to remain above historic minimum levels.

In the uncoordinated scenario, we use the nine-year average from 2012--2021 consumption on a daily resolution to model the annual gas consumption. 
We split the daily consumption curve into a constant base load and a variable seasonal load. We rescale the variable load according to the reductions assumed for heating and energy production and rescale the base load corresponding to the average reduction in the other uses.
In this paper, we assume that the shortfall of Russian gas starts on June 1, 2022, where the storage level was at 3.63 bcm (33.3\% full) \citep{agsi_gas_storage}. We use the reduced in- and outflow as described in Table~\ref{tab:scen} and set the minimum storage level to 1.46 bcm. This allows a net extraction of 2.11 bcm between June 1, 2022 and May 31, 2023. Our simulation hits the minimum storage level in April 2023, which is well above the historic minimum of 1.35 bcm in March 2022.

To simulate storage levels at the EU level, we rely on monthly data provided by Eurostat's natural gas supply statistics \citep{eurostat_supply_consumption}.
We average over the last two years for consumption and inflow levels. The EU gas storage contained 60.24 bcm (47.7\% of maximum capacity) on June 1, 2022 \citep{agsi_gas_storage}. 
Again, assuming a constant natural gas supply and a rescaled, variable consumption (according to the values in Table~\ref{tab:scen}), we can calculate the monthly storage level. 
We set the minimal gas storage level to 23.7 bcm (18.7\%), which is slightly above the historic minimal gas storage of 22.10 bcm (17.5\%) in March 2018.
This allows a net extraction of 28 bcm from June 1, 2022, to May 31, 2023, resulting in the minimum storage level in March 2023. Assuming a proportional distribution in the EU-cooperation scenario, Austria is allocated 28/412 * 9.34 bcm = 0.64 bcm.

Note that in all these computations, we have assumed that the Austrian storage is only used to provide natural gas for domestic consumption, which is typically not the case. For example, German companies have rented parts of Austrian storage -– a fact that we neglect in the uncoordinated Scenario B. Thus, our estimate of the extent to which storage can be used for mitigating import shortfalls might be optimistic. Note that, in a truly uncoordinated scenario, if Austria would cut off Germany from its rented storage, Germany could retaliate by not allowing Austria to use its pipeline infrastructure to import LNG. This scenario highlights that cooperation among European countries is essential.

We stress further that the effective gas shortage over a one-year horizon crucially depends on the timing of the stop of Russian gas imports. We have analyzed the economic consequences of an import stop on June 1, 2022, which is relatively early in the net injection cycle (minimum gas storage levels typically occur in April and maximum levels in early November), and after a winter with historically low gas storage levels. 
In the meantime, the EU Gas Storage Regulation was passed, introducing an 80\% gas storage target before winter 2022 \citep{storage_regulate2022} and 90\% storage targets for subsequent years. Both the EU and Austria exceeded these targets, filling gas storages to around 95\% of their capacity before winter 2022~\citep{agsi_gas_storage}. 
Thus, in case of a drastic import disruption a few weeks later than in our simulations there would be substantially more leeway to counteract the shortfalls, resulting in significantly smaller adverse economic impacts.

\subsection{Fuel substitution in the power and heat sector} \label{apx:_mit_sub}
Electricity generation and district heating plants are among the largest gas consumers in Austria. Gas is currently the only non-renewable source in Austria's power mix (besides non-renewable electricity imports).
Thus, switching gas power plants to alternative fuels, such as oil, can substantially reduce gas demand in the short run. 
However, in Austria, dispatchable gas power plays a crucial role in grid stabilization, which, in the short term, cannot easily be provided by alternative energy sources. Moreover, gas consumption in the electricity sector is highly seasonal. These plants burn natural gas mostly during the winter months to meet higher electricity demand and to compensate for lower production from renewable sources.

To conservatively estimate the amount of gas needed to stabilize the grid, we assume that only gas-fired power plants, especially open-cycle plants, perform the critical function of grid stabilization. 
We make this assumption based on the fast ramp-up times of open-cycle gas turbines \citep{ocgt_frequency}.
According to the dispatch data from the Austrian Power Grid \citep{apg_statistik}, grid stabilization in 2019 required the consumption of 0.32 TWh of electrical energy.
This corresponds to 0.32 TWh/0.35 = 0.9 TWh, or about 0.09 bcm of gas per year, assuming an efficiency of 35\% for open-cycle gas turbines \citep{ocgt_performance_analysis,ocgt_efficiency}.
We assume that in the short term, it is not feasible to substitute gas-based grid stabilization with alternative sources. Therefore, the electricity sector always requires 0.09 bcm of gas annually for grid stabilization purposes.

Some gas-fired power plants in Austria have the capability to switch between gas and oil, making them potential candidates for fuel switching in the event of a severe gas shortage.
For example, as of 2020, the 800 MW CHP plant W\"armekraftwerk Thei\ss could still switch between oil and gas fuels \citep{evn}. 
Thus, we assume that in times of crisis, gas power capacities amounting to 800 MW of gas can be converted to alternative fuels, although we are agnostic whether this happens due to fuel-switching of a single large plant, such as W\"armekraftwerk Thei\ss, or several smaller plants, or through retrofitting already-retired coal plants (as mandated by the Austrian government; see Footnote~\ref{foot:switch}).

\begin{figure}
    \centering
    \includegraphics[width=0.9\textwidth, trim={0 .7cm 0 0},clip]{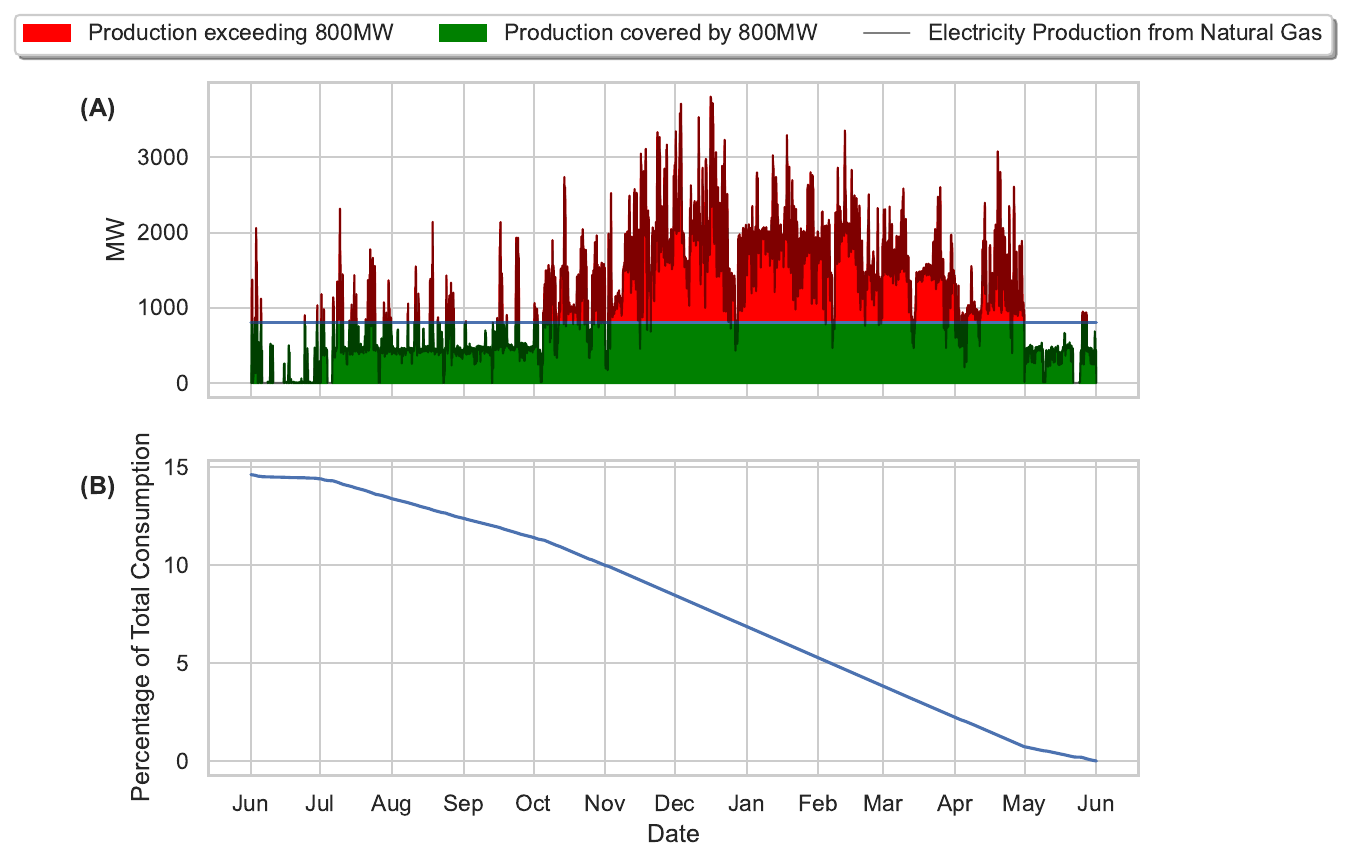}
    \caption{
    \textbf{(A)} {\bf Electricity production from gas} 
   \citep{energy_charts_at}. The horizontal line represents the estimated capacity of 800 MW, which could be provided by other fossil sources (oil or coal). The red area indicates electricity production from gas exceeding 800 MW, which we assume cannot be replaced by alternative fuels. This represents around 40\% of the electricity produced by fossil fuels.
   \textbf{(B)} {\bf Gas savings depending on the timing of substituting 800 MW of gas capacity to alternative fuels.}
Both panels are based on data from June 2021 to June 2022.
}
    \label{fig:apx:elegen}
\end{figure}

To estimate the amount of gas saved by this substitution, we analyzed historical data on electricity generation from natural gas between 2020 and 2021, collected at 15-minute intervals \citep{energy_charts_at}. This data is visualized in Fig.~\ref{fig:apx:elegen}(A). Based on our fuel substitution assumptions, the red area representing electricity production exceeding 800 MW must be provided by gas-fired power stations. The green area below the 800 MW line represents the load that could be met by alternative fuels.
As shown in Fig.~\ref{fig:apx:elegen}(B), the realized gas savings depend on the timing of the fuel substitution.
If the 800 MW gas capacity is switched to alternative fuels in the first summer months, gas savings would amount to about 15\% of annual consumption. 
We assume a 4.5-month time horizon to switch to alternative fuels (mid-October), resulting in 10.5\% of annual gas consumption saved (see Table~\ref{tab:scen}).
This substitution corresponds to a 40\% reduction of gas consumption in the power sector. For comparison, a study in Germany found a reduction potential of 43\% in the power sector's gas consumption~\citep{iek}.

We acknowledge the possibility of practical limitations that may make fuel switching difficult but emphasize its large potential for reducing gas demand. For example, by switching power plants, with a combined capacity of 350 MW, which would be equivalent to Simmering’s Block 3 \citep{wienenergie}, Austria could save an additional 0.40 bcm per year. If such a switch is technically possible, these costs are likely small compared to the economic losses created by gas shortages in industrial sectors. We point out that fuel switching requires the deployment of climate-damaging fuels, counteracting national emission targets. Thus, while fuel switching in the power and heat sectors exhibits large gas-saving potentials in the short term, reducing long-term gas dependencies will crucially depend on the increased deployment of low-carbon energy and storage technologies.

The updated energy balances show that gas consumption in the power and heating sector has declined by 19\% in 2022 compared to 2019--2021 average levels. Note, however, that these observed reductions might be due to various factors, including observed reductions in demand for electricity (e.g., through higher prices) and an increase in renewable energy generation \citep{econtrol-stat23}.

\subsection{Savings in heating} \label{apx:_mit_heat}
Given the drastic scenario of an immediate 80\% gas supply shock, we assume a reduction of gas usage for room heating. These reductions can be driven by price increases but also by other factors, such as public campaigns calling for reducing gas consumption. 
In our analysis, we assume an average room heat reduction of approximately 1$^\circ$C which we expect to translate into a 6\%-reduction of gas demand for heating \citep{palmer2012much}. We assume that the reduction in room heating applies to industry and services, households, and the public sector. As a result, savings in household heating amount to 0.11 bcm, and the savings in industry room heating (before rationing) amount to 0.06 bcm.

With hindsight, our assumption of 6\% heat-related gas savings has been rather conservative. Using the updated energy balance data, we observe that private households have reduced annual gas consumption in 2022 by about 18\% compared to 2019--2021 average levels. However, as pointed out by the Ministry of Climate Action, Environment, Energy, Mobility, Innovation and Technology, reduced gas consumption levels are partially explained by warmer temperatures.
In other EU member states, household gas consumption has declined, too. For example, in the same time period, German households have reduced gas consumption by 12\%, Belgian households by 14\%, Hungarian households by 8\%, and Irish households by 9\%\footnote{
See: \url{https://energie.gv.at/gas/gas} and \url{https://www.bruegel.org/dataset/european-natural-gas-demand-tracker}.
}.

\subsection{Pipelines} \label{apx:_mit_pip}
Natural gas is used as the main energy carrier to operate pipeline infrastructures. The distribution of gas through existing pipelines in Austria required 0.29 bcm in 2021, representing 3.1\% of the total gas consumption (Fig.~\ref{fig:gas-consume-type}A). As consumption falls, the need for pumping declines, and consequently, gas usage for operating the infrastructure is reduced. We have assumed that gas usage for pipeline operation remains a fixed share of total consumption. 
This amounts to 0.05 bcm and 0.11 bcm of savings in Scenarios A and B, respectively.

%% file: appendix/gas_dependency.tex
\section{Estimating the gas dependency of firms}
\label{apx:gasdepend}

To estimate the number of firms that use gas in production, we build on the material input surveys \citep{stat_guetereinsatz} and the structural business statistics \citep{stat_sbs}.
The material input surveys report the annual natural gas consumption for one- to four-digit levels of the following OENACE sectors: B (Mining), C (Manufacturing), D (Energy), E (Utilities), and F (Construction). The total gas consumption in each sector refers to the gas consumed by firms whose annual revenues exceed 10 million EUR and employ at least 20 workers, resulting in 1,388 reports. 
The gas consumption reported in the input material statistics only accounts for 81.3\% of the total industry gas consumption reported in the energy balance \citep{stat_nutzenergie}. Assuming full coverage of the input material surveys, the remaining 18.7\% can be attributed to firms that fall below the reporting threshold of 10 million EUR turnover and 20 employees.

The structural business statistics contain industry-specific information on the total number of firms, as well as some details on employment- and revenue-based firm size distribution. 
We estimate the share of gas-dependent firms in each sector by dividing the number of firms per sector that report the use of gas by the number of firms in the same sector that are, in principle, subject to the reporting obligation. We assume that the same ratio of gas-using to non-gas-using firms also holds for smaller firms that are not obliged to report gas consumption.

The amount of natural gas allocated to industry, as shown in Table~\ref{tab:scen}, contains gas for both production and room heat. 
After taking into account that firms reduce room temperatures by 1$^{\circ}$C (Appendix~\ref{apx:_mit_heat}), we assume that an industry's gas usage for room heat scales proportionally with their level of production. For example, if a sector experiences a 10\% gas supply shock, the reduction in gas supply is allocated proportionally to room heat and production. Note, however, that in our model economic shocks only propagate from firms that use natural gas in production.

%% file: appendix/model.tex
\section{Economic model}
\label{apx:model}

The economic model used in this study has been described in detail in \cite{pichler2022forecasting}. Here, we provide a concise description of the model focusing on the adaptation undertaken to suit the context of a gas supply crisis.
In this model, the economy initially rests in a steady state until it experiences exogenous shocks which in our study correspond to supply input shortages. If a gas-using sector faces shortages in gas inputs, its productive capacity is reduced, and it will need to scale down its production.

The model runs at a daily time resolution, and we denote a time step (a day) by $t$.
There are $N$ sectors (one representative firm for each sector) and one representative household.
Let $x_{i,t}$ denote the total output of sector $i$ at time $t$ and $Z_{ji,t}$ the intermediate consumption by sector $i$ of good $j$. 
Following basic national accounting relationships, the output of $i$ is equal to
\begin{equation} \label{eq:totout}
    x_{i,t} = \sum_{j=1}^N Z_{ij,t} + c_{i,t} + f_{i,t},
\end{equation}
where $c_{i,t}$ is household consumption of good $i$ at time step $t$ and $f_{i,t}$ is all other (exogenous) final demand, including investments, government consumption, and exports. 
We let $l_{i,t}$ denote labor compensation (salary and wages) to workers in sector $i$. Since our model does not distinguish between prices and quantities, $l_{i,t}$ also indicates the number of workers employed in sector $i$. Profits of sector $i$ can then be written as 
\begin{equation} \label{eq:profit}
    \pi_{i,t} = x_{i,t} - \sum_{j=1}^N Z_{ji,t} - l_{i,t} - e_{i,t},
\end{equation}
where $e_{i,t}$ represents all other expenses (taxes, imports, etc.).

\subsection{Demand}
\label{apx:demand}

The total demand faced by sector $i$ at time $t$, $d_{i,t}$, is the sum of the demand from all its customers,
\begin{equation}
    d_{i,t} = \sum_{j=1}^N O_{ij,t} + c^d_{i,t} + f^d_{i,t},
\end{equation}
where $O_{ij,t}$ (for \emph{orders}) denotes the intermediate demand from sector $j$ to sector $i$, $c_{i,t}^d$ represents (final) demand from households and $f_{i,t}^d$ denotes all other final demand (e.g. government or non-domestic customers).

The dynamics of intermediate demand are similar to those in \cite{romanoff1986capacity, henriet2012firm, hallegatte2014modeling, inoue2019firm} and \cite{ reissl2022assessing}. Specifically, the demand from sector $i$ to sector $j$ is
\begin{equation} \label{eq:order_interm}
    O_{ji,t} = A_{ji} d_{i,t-1} + \frac{1}{\tau} \Big( n_i Z_{ji,0} - S_{ji,t-1} \Big).
\end{equation}
Sector $i$ holds the naive expectation that demand on day $t$ will be the same as on day $t-1$, and demands an amount $A_{ji} d_{i,t-1}$ from $j$, where $A$ represents the technical coefficient matrix with elements $A_{ji} = Z_{ji,0}/x_{i,0}$ (note that we use the zero time index to indicate the pre-shock steady state).
The second term in Eq.~\eqref{eq:order_interm} describes intermediate demand induced by desired reduction of inventory gaps. Due to the dynamic nature of the model, demanded inputs cannot be used immediately for production. 
Instead, sectors use an inventory of inputs in production. $S_{ji,t-1}$ denotes the stock of input $j$ held in $i$'s inventory at the beginning of time $t$. Each sector $i$ aims to keep a target inventory $n_i Z_{ji,0}$ of every required input $j$ to ensure production for $n_i$ further days.
The parameter $\tau$ indicates how quickly a sector adjusts its demand due to an inventory gap. Small $\tau$ corresponds to responsive sectors that aim to close inventory gaps quickly. In contrast, if $\tau$ is large, intermediate demand adjusts slowly in response to inventory gaps.

Compared to the model presented in \cite{pichler2022forecasting}, we substantially simplify the dynamics of household demand. We let consumption demand for good $i$ be
\begin{equation}\label{eq:cd}
    c^d_{i,t}= \theta_{i} \; c^d_{0} \; \frac{l_t - l_0}{l_0} , 
\end{equation}
where $\theta_{i}$ is a preference coefficient giving the share of goods from sector $i$ out of initial total consumption demand $c^d_{0} = \sum_i c^d_{i,0}$. We assume that preferences are fixed in time and are given by the initially observed sector-specific consumption shares of households $\theta_{i} = c_{i,0} / \sum_i c_{i,0}.$ $l_t = \sum_i l_{i,t}$ represents total labor income and the term $l_t/l_0 -1$, thus, captures Keynesian income effects. In this simplified representation, a $z$-\% reduction in labor income compared to initial levels results in a $z$-\% reduction in consumption demand. Note that we do not consider shocks to other final demand. Thus, $f_{i,t}^d = f_{i,0}$.

\subsection{Supply}
\label{apx:supply}
Every sector aims to satisfy incoming demand by producing the required amount of output. Production is subject to the following input constraints. First, the productive capacity of a sector that relies on gas inputs in production is constrained by the amount of available gas inputs. Recall from Eq.~\eqref{eq:direct} that $\Delta x_{i,t}^d$ is the direct output shock induced by shortages in gas supply. Thus, we denote the productive capacity of a sector considering gas supply availability by
\begin{equation} \label{eq:xcap_gas}
x_{i,t}^{{G}} = ( 1 + \alpha_i g_i \epsilon^G) \; x_{i,0} \quad \forall i \in \mathcal{G},
\end{equation}
where $\mathcal{G}$ is the set of sectors that require natural gas as production inputs and $\epsilon^G \in [-1,0]$ denotes the gas shock to industrial sectors. 

Second, the productive capacity of a sector also depends on the amount of available labor input. Again, we assume that productive capacity depends linearly on labor inputs,
\begin{equation} \label{eq:xcap}
x_{i,t}^{{l}} =  \frac{l_{i,t}}{l_{i,0}}x_{i,0},
\end{equation}
where we have assumed that every sector produces at full capacity in pre-shock times.

Third, the production of a sector can be constrained due to an insufficient supply of other inputs caused by production network disruptions. 
Intermediate input-based production capacities depend on the availability of inputs in a sector's inventory and its production technology, i.e.
\begin{equation} \label{eq:xinp_general}
x_{i,t}^{\text{inp}} = \mathcal{F}_i( S_{ji,t}, A_{ji} ),
\end{equation}
where $\mathcal{F}$ is the production function. 
Taking all possible production constraints into account, realized production at time step $t$ is 
\begin{equation} \label{eq:realprod}
x_{i,t} = \min \{ x_{i,t}^{G}, \; x_{i,t}^{l}, \; x_{i,t}^{\text{inp}}, \; d_{i,t}  \}.
\end{equation}

To calibrate the function $\mathcal{F}$, \cite{pichler2022forecasting} conducted a survey to identify critical inputs and introduced the partially binding Leontief (PBL) production function. This production function assumes that output reductions of upstream suppliers propagate to downstream customers only if these suppliers provide critical inputs to the production of the downstream customers. In mathematical terms, we have
\begin{equation} \label{eq:xinp_ihs1}
x_{i,t}^{\text{inp}} = 
\min_{j \in \{ \mathcal{V}_i \} } 
\left \{ \;  \frac{ S_{ji,t} }{ A_{ji} } \right \},
\end{equation}
where $\mathcal{V}_i$ is the set of \textit{critical} inputs (see Appendix~\ref{apx:calibration}) to industry $i$. We use this specification as our baseline model setup. 

We further use two extreme specifications of production functions to derive lower and upper bounds of indirect shock amplification. Specifically, we use the Leontief production function to derive an upper bound of shock amplification, in which every positive entry in the technical coefficient matrix $A$ is a binding input to a sector. In this case, we have 
\begin{equation} \label{eq:xinp_leo}
x_{i,t}^{\text{inp}} = 
\min_{ \{ j: \; A_{ji} > 0 \} } 
\left \{ \;  \frac{ S_{ji,t} }{ A_{ji} } \right \}.
\end{equation}

Under the Leontief production function, a sector halts production immediately if inventories of any input run down, even if the input represents a small and potentially negligible share of expenses.
To derive a lower bound of shock amplification, we implement a linear production function, in which all inputs are perfect substitutes. In this case, production in a sector continues even if some inputs cannot be provided, as long as there is a sufficient supply of the other inputs: 
\begin{equation} \label{eq:xinp_linear}
    x_{i,t}^{\text{inp}} =   \frac{ \sum_j S_{ji,t} }{ \sum_j A_{ji} } .     
\end{equation}

We explore the impact of applying alternative specifications of input constraints and heterogeneous inventories in Appendix~\ref{apx:sense}.

\paragraph{Rationing.}
In the presence of production capacity and/or input bottlenecks, sectors' output may be smaller than total demand (i.e., $x_{i,t} < d_{i,t}$), in which case sectors ration their output across customers. We assume simple proportional rationing, although price-based allocation mechanisms and different rationing mechanisms could be considered \citep{pichler2022simultaneous}.
The final delivery from sector $j$ to sector $i$ is the share of orders received,
\begin{equation}
    Z_{ji,t} = O_{ji,t} \frac{x_{j,t}}{d_{j,t}}.
\end{equation}
Households receive a share of their demand
\begin{equation}
    c_{i,t} = c_{i,t}^d \frac{x_{i,t}}{d_{i,t}},
\end{equation}
and the realized final consumption of agents with exogenous final demand is 
\begin{equation}
    f_{i,t} = f_{i,t}^d \frac{x_{i,t}}{d_{i,t}}.
\end{equation}

The inventory of $i$ for every input $j$ is updated according to
\begin{equation}
    S_{ji,t+1} = \max \left\{ S_{ji,t} + Z_{ji,t} - A_{ji} x_{i,t}, 0 \right\}.
\end{equation}

\paragraph{Employment dynamics.}
Firms adjust their labor force depending on which production constraints are binding. If the labor constraint $x_{i,t}^{{l}}$ is binding, sector $i$ decides to try to hire as many workers as necessary to make the capacity constraint no longer binding. Conversely, if gas input constraints $x_{i,t}^{G}$, other intermediate input constraints $x_{i,t}^{\text{inp}}$ or demand constraints $d_{i,t}$ are binding, sector $i$ lays off workers until capacity constraints become binding.  
More formally, at time $t > 0$ labor demand by sector $i$ is given by $l^d_{i,t}=l_{i,t-1}+\Delta l_{i,t}$, with 
\begin{equation}
    \Delta l_{i,t} = \frac{l_{i,0}}{x_{i,0}}\left[ \min\{x_{i,t}^{\text{inp}},d_{i,t}\} - x_{i,t}^{\text{cap}}\right].
\end{equation} 
Following \citep{del2021occupational}, we assume that adjustment of labor inputs is sluggish. Sectors can increase their labor force only by a fraction $\gamma_{\text{H}}$ and decrease their labor force only by a fraction $\gamma_{\text{F}}$ in the direction of their targets.The sector-specific employment evolves according to
\begin{equation} \label{eq:labor_evolution}
    l_{i,t} = 
    \begin{cases} 
    l_{i,t-1} + \gamma_{\text{H}} \Delta l_{i,t} &\mbox{if } \; \Delta l_{i,t} \ge 0, \\
    l_{i,t-1} + \; \gamma_{\text{F}} \Delta l_{i,t} &\mbox{if }  \Delta l_{i,t} < 0.
    \end{cases}
\end{equation}

\subsection{Gas consumption in production} \label{sec:gascons}
In our model, we keep track of realized gas consumption. In case the production of a gas-using sector is limited by other constraining factors (other intermediate inputs, labor, or demand), i.e., $x_{i,t} < x_{i,t}^G$, this sector will use less gas than has been allocated to the sector based on the gas shock scenario considered. We assume that residual gas that industrial consumers have not consumed at time $t$, is available for industrial consumption at the next time step $t+1.$ 

Let $G_{i,t}$ denote the gas consumption of sector $i$ at time step $t$ (in cubic meters). We assume that sector $i$'s gas intensity in production is constant in the short-time period considered and given by observed pre-shock intensity levels $\psi_{i} = G_{i,0} / x_{i,0}$. Thus, if sector $i$ produces the output $x_{i,t}$ according to Eq.~\eqref{eq:realprod}, it consumes $G_{i,t} = \psi_{i} x_{i,t}$ of natural gas. If the gas shock to industry $\epsilon^{G}$ happens at $t=1$, the gas budget available to industry is $B_{t=1}^G = (1-\epsilon^{G}) \sum_i G_{i,0}$, and, by taking realized gas consumption into account, evolves according to $B_{t+1}^G = B_{t}^G - \sum_i G_{i,t}$.

\subsection{Calibration and initialization} \label{apx:calibration}
In Table~\ref{tab:econmodelpars}, we give an overview of the model parametrization. Most of the model variables and parameters can be inferred directly from publicly available data. For example, sectoral economic variables have been taken from national input-output accounts. 
We mostly follow the parametrization as provided in \cite{pichler2022forecasting} for some parameters that cannot be estimated directly from available data.

For example, since we did not have access to Austria-specific inventory data at the sector level, we have initialized the inventory target parameter $n_i$ with the provided inventory data from the United Kingdom. Since inventory levels play an important role in buffering the downstream propagation of shocks, we investigate the role of varying these parameters in the sensitivity analysis (Appendix~\ref{apx:sense}).
The inventory adjustment timing $\tau$ represents a free parameter that could be calibrated from historical data. However, we show in the sensitivity analysis that varying this parameter has limited impacts on our results and set $\tau=10$, implying that firms aim at filling their inventory gaps within ten days.
The parameters $\gamma_H$ and $\gamma_F$ control the rate of hiring and firing. We choose $\gamma_H = 1/30$ and $\gamma_H = 1/15$, yielding rather rapid adjustment of the labor force, with firms adjusting to their demanded labor in about a month when hiring and 15 days when firing. Note that this calibration choice implies that firing happens faster than hiring, which we consider reasonable.

Variables and parameters relating to estimating direct economic shocks from gas supply shortages have been discussed at length in Sections~\ref{sec:gasscenario} and \ref{sec:directshock}. The share of firms in a given sector that uses gas in production, $g_i$, has been estimated from data that can be acquired from Statistik Austria (material input surveys and structural business statistics). While we have focused on two specific gas supply shortage scenarios in the main text, we have modeled economic impacts across the whole spectrum of possible sectoral gas shock values $\epsilon^G_i$. Note that in all our simulations, we have assumed that gas shocks are identical across sectors, i.e., $\epsilon^G_i = \epsilon^G$.
A key parameter in our analysis is $\alpha_i$, specifying the output-dependency of gas-using firms of sector $i$. Without better information, we have adopted the rather conservative assumption of $\alpha_i=1$. We emphasize that this choice matters, as we demonstrate in more detail in the sensitivity analyses of Appendix~\ref{apx:sense}.

\begin{table}[pos=htbp]
	\centering
		\begin{tabular}{|l|c|c|}
			 \hline
			 Model setup and parameters & Symbol & Value \\
			 \hline
 			 Number of sectors & $N$ & 63 \\
			 Input-output variables & $Z_{ij}$, $A_{ij}$, $f_{i}$, $x_{i}$, $l_{i}$, $c_{i}$, $\theta_i$ & IO tables (Statistik Austria) \\
			 Inventory targets & $n_i$ & as in  Pichler et al. (2022) \\
          Critical inputs & $\mathcal{V}_i$ & as in  Pichler et al. (2022) \\  
			 Inventory adjustment & $\tau$ &10 \\
			 Production function & $\mathcal{F}$ & linear, PBL, Leontief \\
			 Upward labor adjustment & $\gamma_H$ & 1/30 \\
			 Downward labor adjustment & $\gamma_F$ & 1/15 \\
    		 Output dependency of gas-using firms & $\alpha_i$ & 1 \\
          Share of firms using gas& $g_i$ & estimated (data from Statistik Austria) \\
    		 Gas shock& $\epsilon^G$ & estimated, scenario-dependent \\
			\hline
		\end{tabular}
	\caption{ \textbf{Overview of model setup and parameters.}  
	}
	\label{tab:econmodelpars}
\end{table}

%% file: appendix/add_results.tex
\section{Model sensitivity analysis}
\label{apx:sense}

\begin{figure}[!ht]
    \centering
    \includegraphics[width=.495\textwidth]{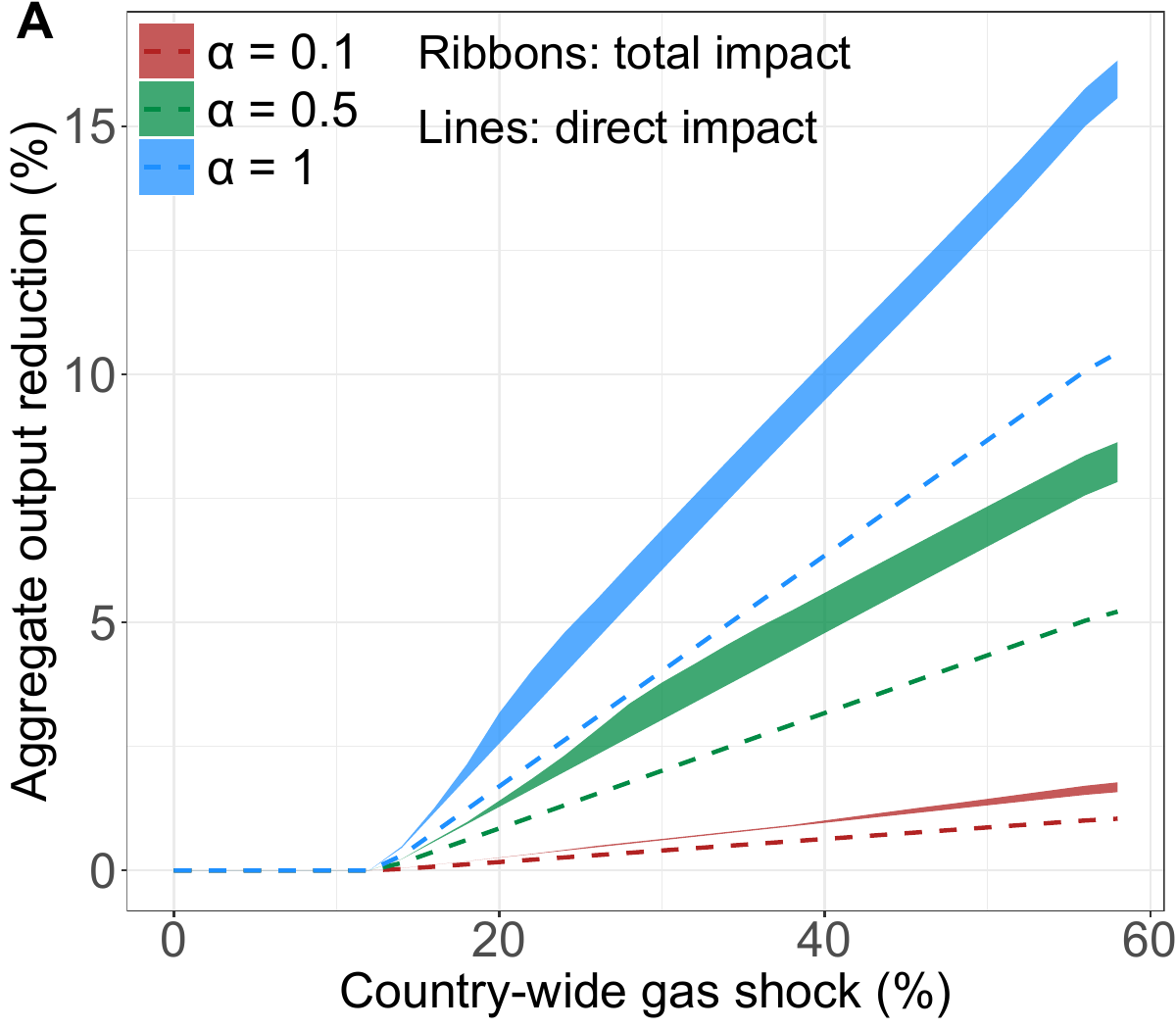}
        \includegraphics[width=.495\textwidth]{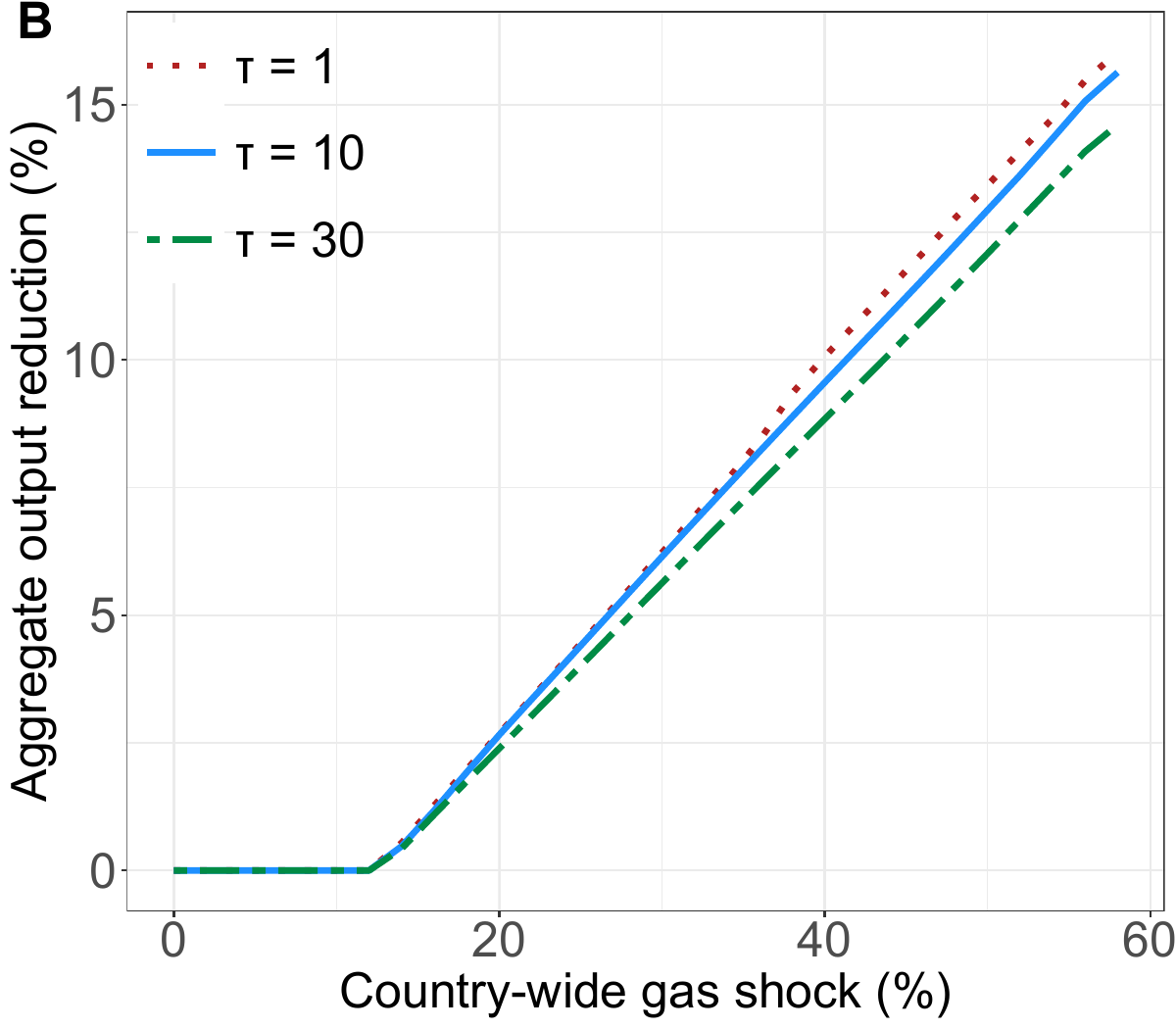}
    \caption{
    {\bf A: The impact of parameter $\alpha$ on economic impact estimates.}
The y-axis shows aggregate output reductions and the x-axis shows the country-wide gas shock. The dashed lines show direct shock estimates and the ribbons reflect total (direct + indirect) impacts based on differing production function assumptions, where the linear production function yields the lower bound and the Leontief production function the upper bound. Baseline estimates of the main text are based on $\alpha = 1$.
{\bf B: The impact of parameter $\tau$ on economic impact estimates.} $\tau=10$ refers to the baseline specification used in the main text.
    }
    \label{fig:alpha}
        \includegraphics[width=0.495\textwidth]{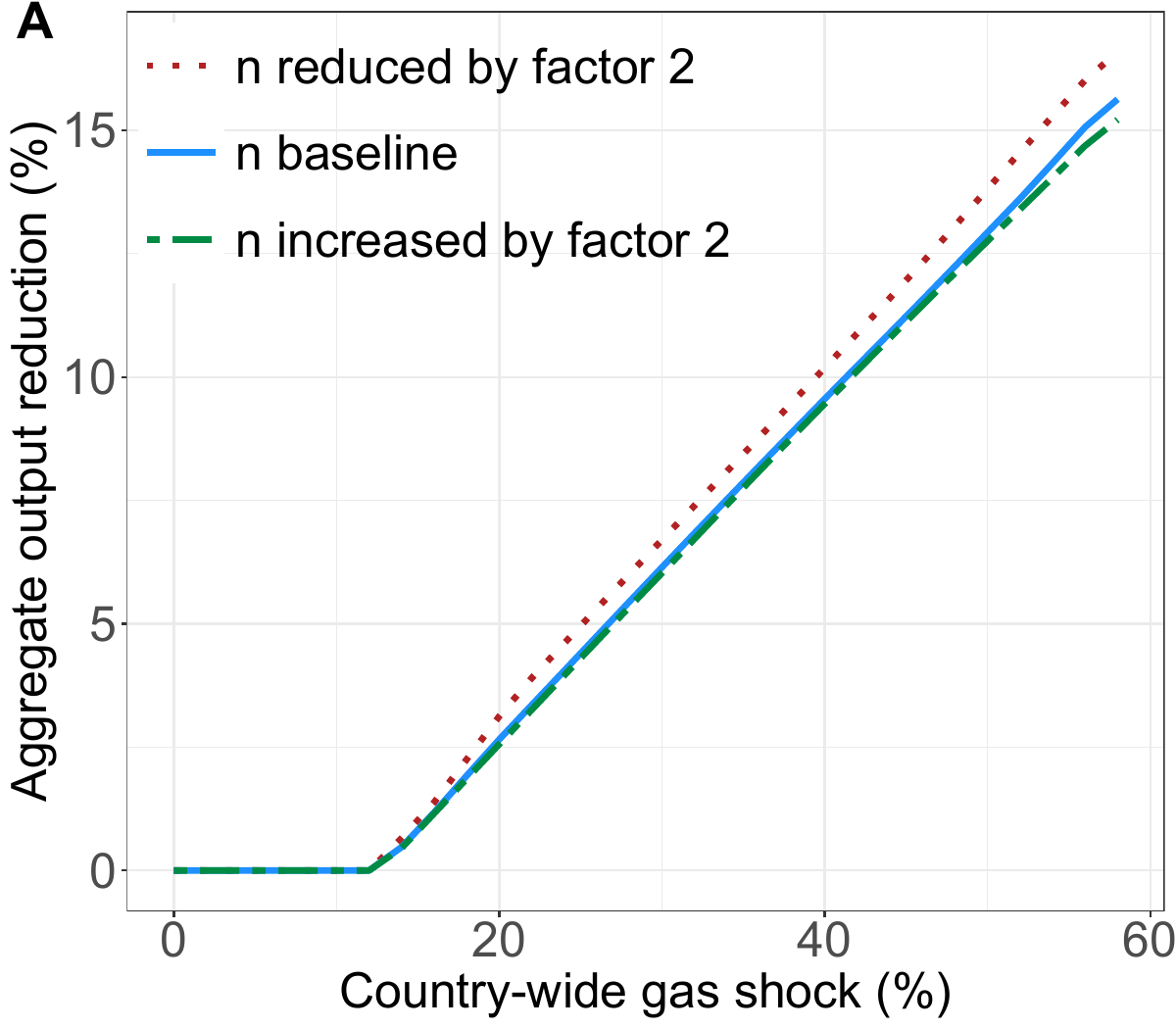}
    \includegraphics[width=0.495\textwidth]{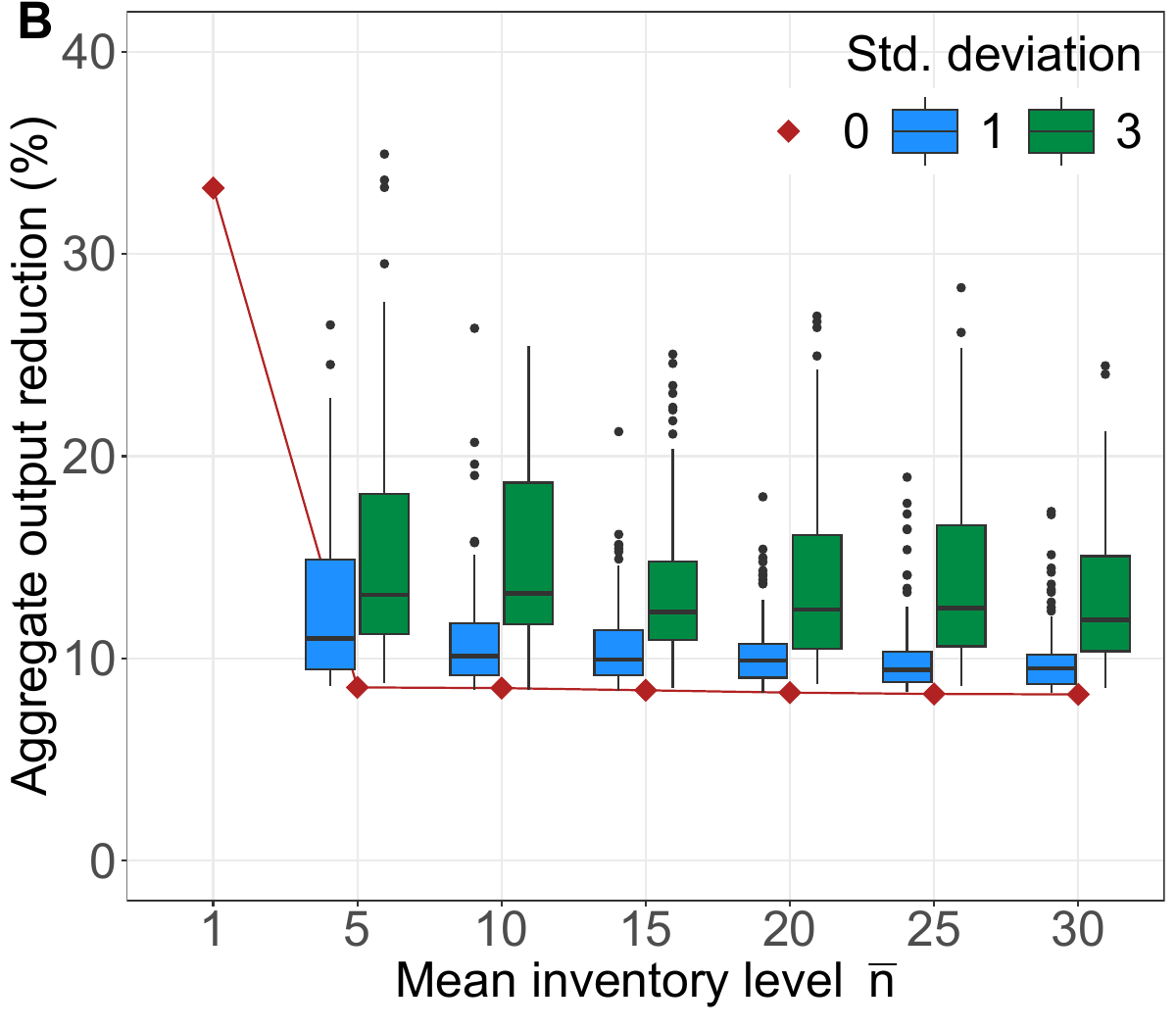}
    \caption{
    {\bf A: The impact of increasing/decreasing inventory levels on economic impact estimates.}
\emph{n baseline} indicates the parametrization of the inventory parameters $n_i$ as indicated in the main text and Appendix~\ref{apx:calibration}. \emph{n reduced by factor 2} uses inventory levels that are only half the size compared to the baseline and \emph{n increased by factor 2} inventory levels that are twice the size compared to the baseline. Note that we have set $n_i^\text{reduced} = \max\{1, 1/2 \; n_i\}$ to ensure that the model starts in the steady state, as $n_i^\text{reduced} < 1 $ would immediately lead to a decline in economic output.
{\bf B: The impact of inventory heterogeneity on economic impact estimates.}
The x-axis shows mean inventory levels $\bar n$. Inventory levels have been generated through drawing from a normal distribution $n_i \sim N(\bar n, \text{S} \bar n)$. We have truncated the distribution such that $n_i \ge 1$. The coloring of the boxplots indicates the standard deviation scaling factor (S). Boxplots are based on aggregate output reductions from running 100 simulations per S. The red diamond (S=0) represents the case of homogeneous inventory levels $n_i = \bar n$. Shocks are based on the Uncoordinated Scenario B.
    }
    \label{fig:inv}
\end{figure}

To better understand the model behavior, we conduct sensitivity analyses by investigating how aggregate output estimates are affected under alternative modeling assumptions. 
A key parameter in our analysis is $\alpha_i$, the output dependence of gas-using firms with respect to gas inputs. We show in Fig.~\ref{fig:alpha}A how economic output depends on the specification of $\alpha$. The figure illustrates that the choice of $\alpha$ strongly affects direct (lines) and total (ribbons) economic impacts. From Eq.~\eqref{eq:direct} the scaling of direct shocks with parameter $\alpha$ becomes evident. Decreasing $\alpha$ by a factor of two also reduces direct economic shocks by a factor of two. When considering the linear production function specification of the model, results for total impact estimates scale almost identical with $\alpha$ as observed for direct impacts. The scaling is somewhat different in the Leontief specification due to nonlinearities introduced in downstream shock propagation effects.

We next investigate how estimated output impacts depend on inventory adjustment speed $\tau$ (Fig.~\ref{fig:alpha}B).
We find that our results are relatively insensitive to changes in $\tau$. Increasing the baseline parametrization ($\tau=10$) by a factor of three results in slightly milder aggregate output reductions. Decreasing this value to $\tau=1$, results in somewhat higher aggregate output reductions.

In Fig.~\ref{fig:inv} we analyze how inventory levels affect total output impacts. In Panel A, we see that results are relatively robust to increasing and decreasing baseline inventory levels by a factor of two. Since we observe only limited downstream shock propagation, increasing inventory levels further does not significantly reduce adverse impacts. As expected, halving inventory levels leads to higher estimated output reductions, although shock amplification effects are still limited.

We next investigate how inventory level heterogeneity affects our results. 
To obtain heterogeneous sectoral inventory levels, we draw them from a normal distribution $n_i \sim N(\bar n, \text{S} \bar n)$ which we left-truncate at $n_i = 1$ to ensure that the model always starts in the steady state.  We then scale the factor $\text{S}$, to see how changing the level of heterogeneity in inventory holdings affects our results. Note that $\text{S}=0$ corresponds to homogeneous inventory holdings where each sector holds an inventory of $n_i = \bar n$ days of production.
As shown in Fig.~\ref{fig:inv}B, adverse economic impacts are extremely severe if every sector holds input inventories of only one day of production. If inventory levels are increased to five days of production for each sector, predicted adverse impacts decline substantially, while further increasing inventory levels only mildly reduces impacts. 
Interestingly, we observe substantial effects of inventory heterogeneity on model predictions. If inventory levels vary by one standard deviation of the mean value, median impacts increase and we find various ``unlucky'' realizations of sectoral inventory levels that result in extremely adverse outcomes. If we increase the heterogeneity further to three standard deviations of the mean, we observe a further significant rise in negative economic impacts. Thus, inventory level heterogeneity plays an important role in shock amplification: more heterogeneous inventories tend to result in larger adverse impacts.